\documentclass[useAMS,twocolumn,trackchanges]{aastex63}

\usepackage{amsmath,amssymb}
\usepackage{natbib}
\usepackage{xspace}

\newcommand{\simgt}{\lower.5ex\hbox{$\; \buildrel > \over \sim \;$}}
\newcommand{\simlt}{\lower.5ex\hbox{$\; \buildrel < \over \sim \;$}}

\newcommand{\myemail}{masato.kobayashi@nagoya-u.jp, kobayashi@ph1.uni-koeln.de}

\newcommand{\m}{${\rm M}_{\rm 200b}$ }

\newcommand{\eg}{e.g.\xspace}
\newcommand{\cf}{c.f.\xspace}
\newcommand{\ie}{i.e.\xspace}
\newcommand{\etc}{etc.}

\newcommand{\msun}{\mbox{${\rm M_{\odot}}$}\xspace}

\newcommand{\zsun}{\mbox{${\rm Z_{\odot}}$}\xspace}

\newcommand{\kms}{{\rm km \, s^{-1}}}

\newcommand{\cmkk}{\mbox{${\rm cm^{-3}}$}}

\newcommand{\myCooL}{\mathcal{L}}

\newcommand{\mymgas}{m_{\rm gas}}
\newcommand{\myboltz}{k_{\rm B}^{}}

\newcommand{\unitk}{\mathbf{\hat{k}}}

\received{--}
\revised{--}
\accepted{--}
\submitjournal{ApJ}

\begin{document}


    \title{Metallicity Dependence of Molecular Cloud Hierarchical Structure at Early Evolutionary Stages}
    \shorttitle{CLOUD FOMRATION IN THE LOW-METALLICITY ENVIRONMENT}
    \shortauthors{Masato I.N. Kobayashi}




    \correspondingauthor{Masato I.N. Kobayashi}
    \email{\myemail}
    \author[0000-0003-3990-1204]{Masato I.N. Kobayashi}
    \affiliation{Division of Science, National Astronomical Observatory of Japan, Osawa 2-21-1, Mitaka, Tokyo 181-8588, Japan}\\
    \affiliation{I. Physikalisches Institut, Universität zu Köln, Zülpicher Str 77, D-50937 Köln, Germany}\\
    \author[0000-0002-2707-7548]{Kazunari Iwasaki}
    \affiliation{Center for Computational Astrophysics, National Astronomical Observatory of Japan, Mitaka, Tokyo 181-8588, Japan}\\
    \author[0000-0001-8105-8113]{Kengo Tomida}
    \affiliation{Astronomical Institute, Graduate School of Science, Tohoku University, Aoba, Sendai, Miyagi 980-8578, Japan}\\
    \author[0000-0002-7935-8771]{Tsuyoshi Inoue}
    \affiliation{Department of Physics, Konan University, Okamoto 8-9-1, Kobe, Japan} \\
    \author{Kazuyuki Omukai}
    \affiliation{Astronomical Institute, Graduate School of Science, Tohoku University, Aoba, Sendai 980-8578, Japan}
    \author[0000-0002-2062-1600]{Kazuki Tokuda}
    \affiliation{Department of Earth and Planetary Sciences, Faculty of Sciences, Kyushu University, Nishi-ku, Fukuoka 819-0395, Japan}
    \affiliation{ALMA Project, National Astronomical Observatory of Japan, Mitaka, Tokyo 181-8588, Japan}
    \affiliation{Department of Physics, Graduate School of Science, Osaka Metropolitan University, 1-1 Gakuen-cho, Naka-ku, Sakai, Osaka 599-8531, Japan}
   


\begin{abstract}
The formation of molecular clouds out of H{\sc i} gas 
is the first step toward star formation.
Its metallicity dependence plays a key role to determine star formation 
through the cosmic history.
Previous theoretical studies with detailed chemical networks 
calculate thermal equilibrium states
and/or thermal evolution under one-zone collapsing background.
The molecular cloud formation in reality, however, involves supersonic flows,
and thus resolving the cloud internal turbulence/density structure in three dimension is still essential.
We here perform magnetohydrodynamics simulations of $20\,\kms$ converging flows of Warm Neutral Medium (WNM)
with 1 $\mu$G mean magnetic field 
in the metallicity range
from the Solar (1.0 $\zsun$) 
to 0.2 $\zsun$ environment. 
The Cold Neutral Medium (CNM) clumps form faster with higher metallicity due to more efficient cooling.
Meanwhile, their mass functions commonly follow $dn/dm\propto m^{-1.7}$ at 
three cooling times regardless of the metallicity.
Their total turbulence power also commonly shows the Kolmogorov spectrum with its 80 percent in 
the solenoidal mode, 
while the CNM volume alone indicates the transition towards the Larson's law.
These similarities measured at the same time in the unit of the cooling time
suggest that the molecular cloud formation directly from 
the WNM alone requires a longer physical time 
in a lower metallicity environment in the 1.0--0.2 $\zsun$ range.
To explain the rapid formation of molecular clouds
and subsequent massive star formation 
possibly within $\lesssim 10$ Myr
as observed in the Large/Small Magellanic Clouds (LMC/SMC),
the H{\sc i} gas already contains CNM volume instead of pure WNM.
\end{abstract}






\section{Introduction}
\label{sec:intro}
Molecular clouds host star formation and thus their formation and evolution
is an essential step for star formation in galaxies and galaxy evolution \citep{Kennicutt2012}. 
The mass and volume of galactic disks are dominated by H{\sc i} gas \citep{Kalberla2009},
but the spatial distribution of star formation rate is correlated more with molecular clouds (traced by CO lines)
rather than H{\sc i} gas \citep{Schruba2011,Izumi2022}.
Therefore, the formation efficiency of molecular clouds out of H{\sc i} gas 
and the resultant molecular cloud properties set the initial condition of galactic star formation.
In particular, 
the metallicity 
controls the heating/cooling rate 
and the thermal state of the interstellar medium (ISM)
\citep{Field1969,Wolfire1995,Liszt2002,Wolfire2003,Glover2014,Bialy2019}
as well as it changes the formation/destruction rate of molecules and the resultant chemical state of the ISM 
(\eg, see \cite{Glover2012} for the CO-to-H$_2$ conversion factor 
and see \cite{Glover2014,Bialy2019} for the conditions of the metallicity and the interstellar radiation field
that makes the H$_2$ cooling and heating important.)
Observations show that the metallicity increases with the cosmic time 
by the metal production from massive stars \citep[\eg,][]{Lehner2016}.
The metallicity also has the galactocentric gradient even within individual galaxies \citep[\eg, the Milky Way][]{FernandezMartin2017}.
Therefore, 
it is crucial to investigate 
metallicity dependence of the ISM evolution below 
the Solar value ($\zsun$)
for the understanding of galactic star formation and cosmic star formation history.

In $1.0\,\zsun$ environments, 
theoretical studies show that
the thermal instability in the H{\sc i} phase initiates the phase transition from 
the Warm Neutral Medium (WNM) to the Cold Neutral Medium (CNM) \citep{Field1969}.
Observational studies of the emission and absorption of
H{\sc i}, CII and CO lines 
show the existence of such multiphase ISM
and they now try to constrain the geometrical structure of 
the WNM and the CNM:
for example, the studies  
with Arecibo Telescope \citep[\eg,][]{Heiles2003b},
with \textit{Hubble} Space Telescope \citep[\eg,][]{Jenkins2011},
with Giant Metrewave Radio Telescope \citep[\eg,][]{Roy2013a},
with \textit{Herschel}, Very Large Array, and IRAM 30m telescopes \citep[\eg,][]{HerreraCamus2017,Murray2018}.
Such measurements are performed also toward lower metallicity environments such as the Large Maggelanic Cloud (LMC)
with Australian Telescope Compact Array \citep[\eg,][]{Marx-Zimmer2000}.
Previous one-zone theoretical studies have developed detailed chemical networks
to comprehensively study the metallicity dependence of the ISM evolution from the present-day to primordial gas,
such as thermal evolution of collapsing protostellar clouds
\citep{Omukai2000,Omukai2005} and the thermal/chemical steady states \citep{Bialy2019}.
Theses studies show that the thermal instability
still plays an important role even in low-metallcity environments with $\gtrsim 10^{-2}\, \zsun$
where fine structure lines of 
[OI] ($63.2\, \mu$m)
and [CII] ($157.7\, \mu$m) are the dominant coolant\footnote{See also
\citealt{Bialy2019} and \citealt{Inoue2015} for the effect of the H$_2$ cooling and the H$_2$ dissociation by UV radiation,
as well as \citealt{Omukai2005} and \citealt{Chon2022} for the heating by the Cosmic Microwave Background radiation at high-redshifts
and \citealt{Omukai2001} for the atomic line cooling in the first star formation).}.

The WNM and the CNM are in the pressure equilibrium under a typical Galactic pressure and metallicity \citep{Field1969,Wolfire1995,Liszt2002,Wolfire2003,Bialy2019}.
Supersonic flows are believed to be an important first step to initiate the thermal instability 
by compressing/destabilizing the previously stable WNM to the thermally unstable neutral medium (UNM), 
which subsequently evolves to the CNM due to cooling.
Such supersonic flows originate from the passage of galactic spiral arms,
the expansion of supernova remnants and H{\sc ii} regions \etc \citep{Inutsuka2015}.
Many authors investigated this condition by performing numerical simulations of WNM converging flows,
initially in one-dimension \citep[\eg,][]{Hennebelle1999,Koyama2000,vazquezsemadeni2006},
and later in multi-dimension \citep[\eg,][]{Koyama2002,Audit2005,Heitsch2006b}.
They show that the dynamically condensing motion of the UNM due to the thermal instability 
results in the formation of turbulent clumpy CNM structures, 
which are important progenitors
of the filamentary structures observed in molecular clouds.

This multiphase nature of the ISM seems ubiquitous also in a wide range of low-metallcity environments
as suggested by large-scale simulations on a galactic/cosmological scale
(\eg, supersonic flows by supernovae, galaxy mergers \citealt{Arata2018},
gas accretion from the host dark matter halo \citealt{Dekel2006}).
Therefore, multi-dimensional numerical studies on low-metallicity molecular clouds 
are still essential 
to understand metallicity dependence of their formation and subsequent star formation,
where the thermal instability and turbulence operate simultaneously.
\cite{Inoue2015} performed three-dimensional simulations of converging flows
as well as a linear stability analysis. They confirmed the development of the thermal instability
as long as the dominance of metal lines in the cooling process under 
modest FUV background with $\log(G_0)>-3$
\citep[\cf,][for the validness of the chemical equilibrium assumption in low-metallicity environments]{Hu2021}.
These previous approaches, however, employ a coarser spatial resolution in a lower metallicity 
aiming at resolving the thermal instability with the same number of numerical cells between different metallicities,
because the maximum growth scale of the thermal instability is larger at lower metallicities.
The CNM clumps on sub-pc scales are not fully resolved yet and their properties and statistics 
in low-metallicity environments
remain unclear.

Recent observations with the Atacama Large Millimeter/submillimeter Array (ALMA)
reveal the existence of filamentary structures whose width is $\sim 0.1$ pc
in the outer disk of the Milky Way and the Magellanic Clouds
(\citealt{Tokuda2019,Fukui2019,Indebetouw2020,Wong2022}, Izumi et al., in prep.),
which is similar to the ones in the Solar neighborhood
\citep{Andre2010,Arzoumanian2011}.
These structures are likely inherited from clumpy/filamentary structures in H{\sc i} phase \citep[see][for 
the comparison of H{\sc i} and molecular gas structures]{Fujii2021}.
Such high-resolution observations from radio to optical/near-infrared bands
toward extragalactic sources will advance significantly 
in the upcoming years by 
ALMA,
James Webb Space Telescope (JWST),
the next generation VLA (ngVLA), 
the Five-hundred-meter Aperture Spherical radio Telescope (FAST),
the Square Kilometre Array (SKA) \etc.
Therefore, understanding of sub-pc scale structures during the molecular cloud formation from H{\sc i} gas
is critical to understand the possible universality of star formation process
in different metallicity environments, \eg, the common existence of filamentary molecular clouds
as observations suggest.

In this article, we perform magnetohydrodynamics simulations of WNM converging flows
to investigate the metallicity dependence of the molecular cloud formation,
especially focusing on the thermal instability development 
and the resultant CNM properties.
We investigate three cases with the metallicities of 1.0, 0.5, 0.2 $\zsun$,
which correspond to the typical values of the Milky Way, LMC, and Small Magellanic Cloud (SMC), respectively. 
By aiming at coherently resolving the turbulence/density structures comparable to the scale of molecular filaments/cores,
we employ the 0.02 pc spatial resolution at all metallicities, 
which is enough to resolve the cooling length of the UNM evolving to the CNM \citep[see][]{Kobayashi2020}.
The typical cooling length of the WNM and UNM is 1 pc (1$\zsun$) -- 3 pc (0.2$\zsun$)
and that of the CNM is 0.1 pc (1$\zsun$) -- 0.3 pc (0.2$\zsun$) in our simulation.
This spatial resolution motivated by recent observations
is higher than previous studies \citep[\eg, compared with][at $0.2\,\zsun$]{Inoue2015},
which enables us to coherently compare the statistics of CNM structures and discuss their possible universality/diversity
between different metallicities.

The rest of this article is organized as follows.
In Section~\ref{sec:method}, we explain our simulation setups,
and show the main results in Section~\ref{sec:main}.
In Section~\ref{sec:imp}, we explain the implications and discussions on the low-metallicity cloud formation 
based on our results.
We summarize our results in Section~\ref{sec:summary}
followed with future prospects.

\section{Method}
\label{sec:method}

\subsection{Basic Equations and Setups}
\label{subsec:code}
To investigate the development of the thermal instability and the formation of 
the multiphase ISM from the WNM,  
we calculate supersonic WNM converging flows by solving the following basic equations: 
\begin{eqnarray}
    &&\frac{\partial \rho}{\partial t} + \nabla_{i}^{} (\rho v_{i}) = 0 \,, \label{eq:CoM}\\
    &&\frac{\partial (\rho v_{i})}{\partial t} 
       + \nabla_{j} (T_{ij} + \rho v_{j}v_{i}) = -\rho\nabla_{i}\Phi \,, \label{eq:EoM}\\
    &&T_{ij} = \left( P + \frac{B^2}{8\pi} \right)\delta_{ij} - \frac{B_i B_j}{4\pi}\,, \label{eq:EST}\\
    &&\frac{\partial e}{\partial t} + \nabla_{i} \left[ (e\delta_{ij} + T_{ij} ) v_{j} \right] \nonumber \\
    &&\qquad = \nabla_{i}\left[\kappa(T)\nabla_{i} T\right] -\rho v_{i} \nabla_{i} \Phi -\rho \myCooL(T,Z)\,, \label{eq:EE}\\
    &&\frac{\partial B_i}{\partial t} + \nabla_{j} \left(v_{j}B_{i} - v_{i}B_{j} \right) = 0 \label{eq:IE}\,,\\
    &&\nabla^2 \Phi = 4\pi{\rm G} \rho \label{eq:Poisson}\,,
\end{eqnarray} 
where $\rho$ is the mass density, $v$ represents the velocity,
$P$ represents the thermal pressure, $T$ without any subscript is the temperature, 
$\Phi$ is the gravitational potential,
${\rm G}$ is the gravitational constant, and 
$\nabla_{i} = \partial/\partial x_i$,
where $x_i$ spans $x, y,$ and $z$.
$\delta_{ij}$ is the identity matrix.
We calculate 
the total energy density,
$e$, as $e = P/(\gamma-1) + \rho v^2/2 + B^2/8\pi$
where the ratio of the specific heat is $\gamma=5/3$.
We introduce the thermal conductivity, $\kappa$, 
as $\kappa(T) = 2.5 \times 10^3 \, T^{0.5} \, 
\mathrm{erg \, cm^{-1} \, s^{-1} \, K^{-1}}$,
which considers collision
between hydrogen atoms \citep{Parker1953}.
\begin{figure}
    \centering{\includegraphics[scale=1.00]{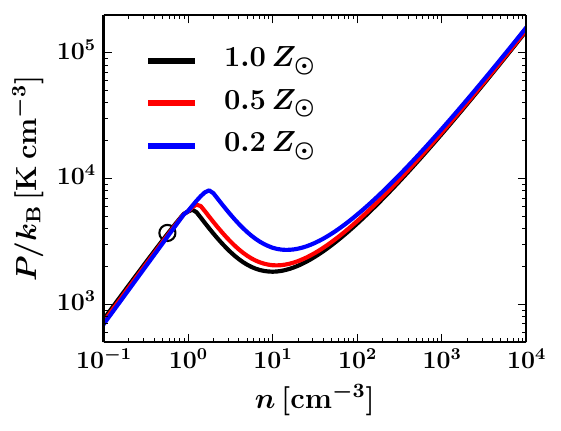}}
    \caption{The thermal equilibrium states as a function of pressure ($P$), density ($n$) and 
             the metallicity (from 1.0 $\zsun$ to 0.2 $\zsun$ in the colors).
             The black circle at the left bottom shows the initial condition at 1.0 $\zsun$.}
    \label{fig:nPkeq}
\end{figure}

$\myCooL(T,Z)$ is the net cooling rate.
We employ the functional form of $\myCooL(T,1.0\,\zsun)$ from \cite{Kobayashi2020} in the case of 
1.0 $\zsun$ condition. This functional form combines the results from
\cite{Koyama2002} in $T \leq 14,577$ K and \cite{Cox1969} and \cite{Dalgarno1972} in $T > 14,577$ K
by considering the cooling rates due to
Ly$\alpha$, C$_{\rm II}$, He, C, O, N, Ne, Si, Fe, and Mg lines
with the photo-electric heating.
The shock heating and compression destabilize the WNM to join the UNM
(defined as $(\partial (\mathcal{L}/T)/\partial T)_P <0$; see \eg,\citealt{Balbus1986,Balbus1995}),
leading to the formation of the CNM clumps.

To investigate the lower metallicity cases in this article, we apply three modifications to $\myCooL(T,1.0\,\zsun)$
to prepare $\myCooL(T,Z)$.
First, the cooling rate due to the metal lines is set to be linearly scaled with the metallicity,
which is a good approximation in $> 10^{-4}\, \zsun$ as long as the metal lines dominate the 
cooling \citep{Inoue2015}.
Second, we set the photo-electric heating rate to be linearly scaled with the metallicity
by assuming that the dust abundance is also scaled with the metallicity.
We, therefore, use $\Gamma_{\rm pe} = 2.0 \times 10^{-26} \left( Z / \zsun \right)$ erg s$^{-1}$. 
Third, we implement the X-ray and cosmic ray heating rates as 
$\Gamma_{\rm X} = 2.0 \times 10^{-27}$ erg s$^{-1}$ and
$\Gamma_{\rm CR} = 8.0 \times 10^{-28}$ erg s$^{-1}$ respectively
\citep{Koyama2000}.
The X-ray and cosmic ray heating processes are subdominant in the metallicity range of this study; 
for example, X-ray (CR) contributes to the 10 \% (30\%) of the total heating rates in the $0.2\, \zsun$ environment.
Their relative importance decreases even more when the metallicity increases 
because the photoelectirc heating rate increases with metallicity. 
We show the detailed functional form of this revised $\myCooL(T)$ in Appendix~\ref{sec:hcr}.
Figure~\ref{fig:nPkeq} shows the thermal equilibrium state, \ie, $\myCooL(T,Z)=0$.
This shows that, in lower metallicity environments, 
the combination of the inefficient cooling and metallicity-independent X-ray and cosmic ray heatings
allows the existence of the WNM phase until higher density
in the range of $1$--$10\,\cmkk$.
In each calculation, we assume that the uniform metallicity distribution in space
as a representation of low-metallicity environments.

We use the publicly available magnetohydrodynamics (MHD) simulation code {Athena++} \citep[][]{Stone2020} 
to solve the basic equations, where we employ the HLLD MHD Riemann solver \citep{Miyoshi2005}
and the constrained transport method to integrate the magnetic fields \citep{Evans1988,Gardiner2005,Gardiner2008}. 
The self-gravity is calculated with the full multigrid method (Tomida et al., in prep.).
\begin{figure}
    \centering{\includegraphics[width = 1.00\linewidth]{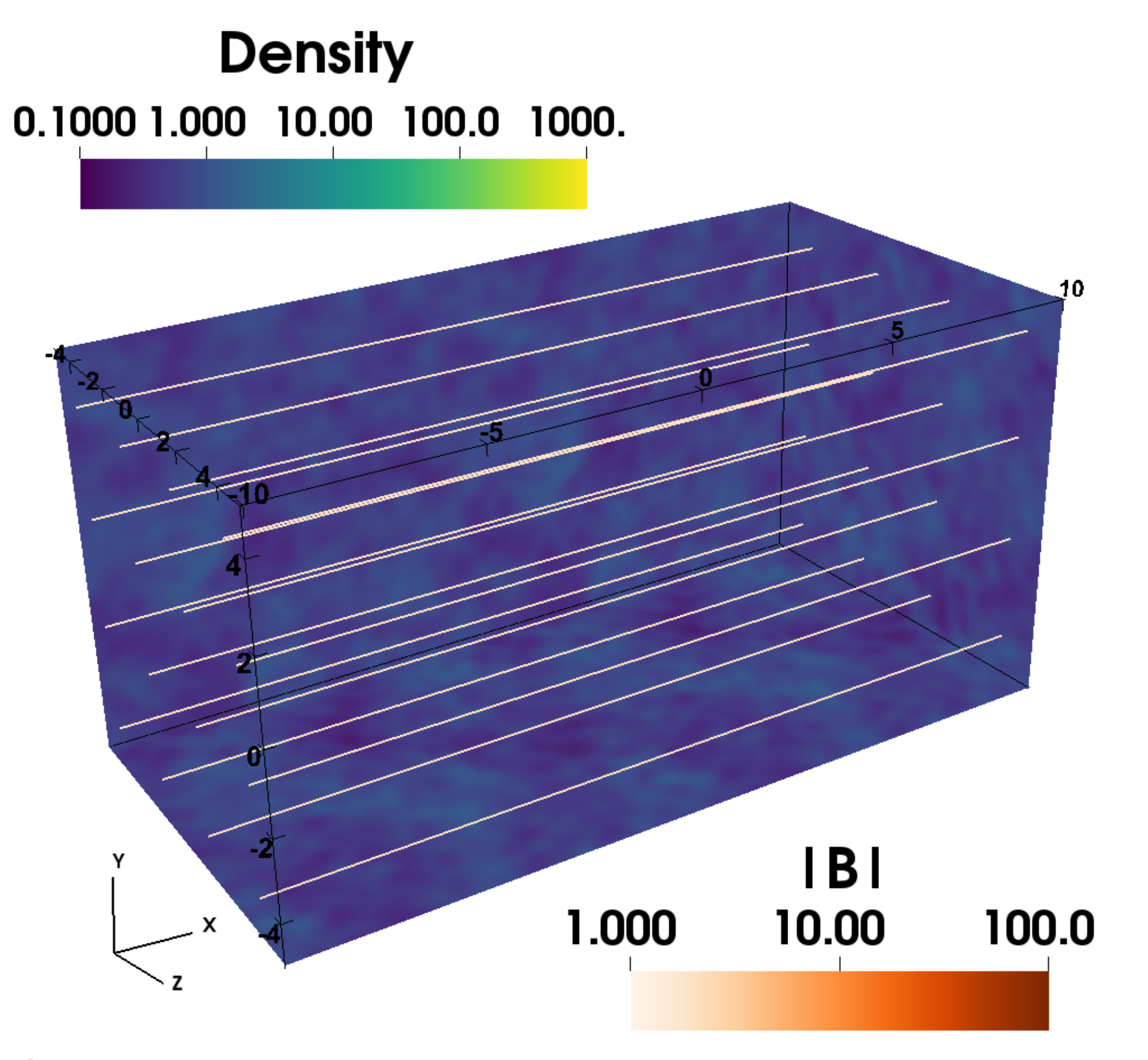}}
    \caption{The initial condition.
             The color shows $n\, [\mathrm{cm}^{-3}]$.
             The white lines represent the initial uniform magnetic fields with the 1 $\mu$G strength. 
             The two supersonic flows are injected through the two $x$ boundaries at $x=-10,10$ pc,
             where the origin of the coordinate is at the box center.}
    \label{fig:init3D}
\end{figure}

Our simulation domain has the size
of $L_{x,y,z}=20,10,10$ pc on each side in the Cartesian coordinate.
We continuously inject supersonic WNM flows from the two $x$ boundaries
so that the two flows collide head-on.
We employ the periodic boundary condition on the $y$ and $z$ boundaries.
The collision forms a shock-compressed layer sandwiched by two shock fronts,
in which the thermal instability converts the WNM to CNM.
We employ $V_{\rm inflow} = 20\,\kms$ as the flow velocity;
this choice corresponds to a representation 
of the late phase of supernova remnant expansion
or normal component of galactic spiral shocks.
The initial velocity field is set as $v_{x} (x) = V_{\rm inflow} 
\tanh (-x/0.78\,\mathrm{pc})$ and $v_{y,z}=0\,\kms$.

The initial WNM flow, and also the injected WNM flow, are thermally stable
with the mean number density $n_0=0.57 \,\cmkk$.
The corresponding pressure, temperature, and sound speed at each metallicity are listed on Table~\ref{table:ParamsMet},
where we use $\rho_0 = n_0 \mu_{\rm M} m_{\rm p}$ with the mean molecular weight $\mu_{\rm M}$ of 1.27 \citep{Inoue2012}.
The ram pressure of the converging flow is $P_{\rm ram}/\myboltz = 3.5 \times 10^4 \, \mathrm{K\,\cmkk}$.
The WNM flows have density fluctuation following the Kolmogorov spectrum 
$P_{\rho}(k)\propto k^{-11/3}$ 
\citep{Kolmogorov1941,Armstrong1995}.
We impose a random phase in each $k$ 
up to $k/2\pi=$ 3.2 pc$^{-1}$, which corresponds to the 0.32 pc wavelength.
The mean dispersion of the density fluctuation is chosen as 
$\sqrt{\langle \delta n_0^2 \rangle}/n_0 = 0.5$.
The injected flows
have periodic distributions smoothly connected to the initial condition,
so that 
we impose 
$\rho (t,x=-10\,\mathrm{pc},y,z) = \rho(t=0,x= 10\,\mathrm{pc} - V_{\rm inflow} t, y, z)$
and 
$\rho (t,x=10\,\mathrm{pc},y,z)  = \rho(t=0,x=-10\,\mathrm{pc} + V_{\rm inflow} t, y, z)$
on the $x$ boundaries where $t$ represents the time.
The interaction between this density inhomogeneity and shock fronts
generate turbulence \citep[\eg,][]{Koyama2002,Inoue2012,CarrollNellenback2014,Kobayashi2022}.

The magnetic field is initially threaded in the $x$ direction with $1$ $\mu$G strength.
Previous studies show that successful molecular cloud formation occurs in
such a configuration where 
the flow and the mean magnetic field 
are close to be parallel \citep{Hennebelle2000,Inoue2008,Iwasaki2019}.
The typical mean field strength 
in H{\sc i} gas varies $1$--$10$ $\mu$G in the Milky Way \citep{Crutcher2012}
and $0.5$--$5$ $\mu$G in the Magellanic Clouds \citep{Gaensler2005,Rainer2013,Livingston2022}.
We, therefore, opt to choose 1 $\mu$G as representative strength in investigating the metallicity dependence.
The detailed dependence on the field strength in low metallicity environments 
is left for future studies at this moment.

We employ the uniform spatial resolution of 0.02 pc to resolve the typical cooling length of the UNM\@.
This resolution is required to have the convergence in the CNM mass fraction after 1.0 $t_{\rm cool}$ (where $t_{\rm cool}$ is the 
cooling time defined as $P/(\gamma-1)/\rho\mathcal{L}$) \citep{Kobayashi2020}.
We identify the shock front position by $P>1.3P_0$ to define the volume of the shock-compressed layer.

Figure~\ref{fig:init3D} shows the three-dimensional view of the initial density field with the uniform 1 $\mu$G magnetic field.
\begin{table}
    \caption{Parameters at different metallicities. $n_0=0.57\,\cmkk$ and $B_0=1\,\mu$G is common at all metallicities.}
    \centering{
        \begin{tabular}{c||c|c|c|c|c}
            \hline
            \hline
              & $P_0/\myboltz$           & $T_0$             & $C_{\rm s,0}$         & $t_{\rm cool}$     & $t_{\rm final}$   \\
              & [$\mathrm{K\,cm^{-3}}$]  & [K]               & [$\kms$]              & [Myr]              & [Myr]             \\ \hline
$1\, \zsun$   & $3.6\times10^3$          & $6.4\times10^3$   & 6.4  & 1.0                & 3.0               \\
$0.5\, \zsun$ & $3.5\times10^3$          & $6.2\times10^3$   & 6.3  & 2.0                & 6.0               \\
$0.2\, \zsun$ & $3.4\times10^3$          & $6.1\times10^3$   & 6.2  & 5.1                & 15.0              \\
            \hline
            \hline
        \end{tabular}
    }\par
    \label{table:ParamsMet}
\end{table}
\begin{figure*}
    \hspace{0.1cm} {\large (a) 1.0 $\zsun$ at $t/t_{\rm cool}=3.0$}
    \hspace{1.2cm} {\large (b) 0.5 $\zsun$ at $t/t_{\rm cool}=1.5$}
    \hspace{1.2cm} {\large (c) 0.2 $\zsun$ at $t/t_{\rm cool}=0.6$} \\
    \begin{minipage}[t]{0.33\textwidth}
        \centering{\includegraphics[width = 1.00\linewidth]{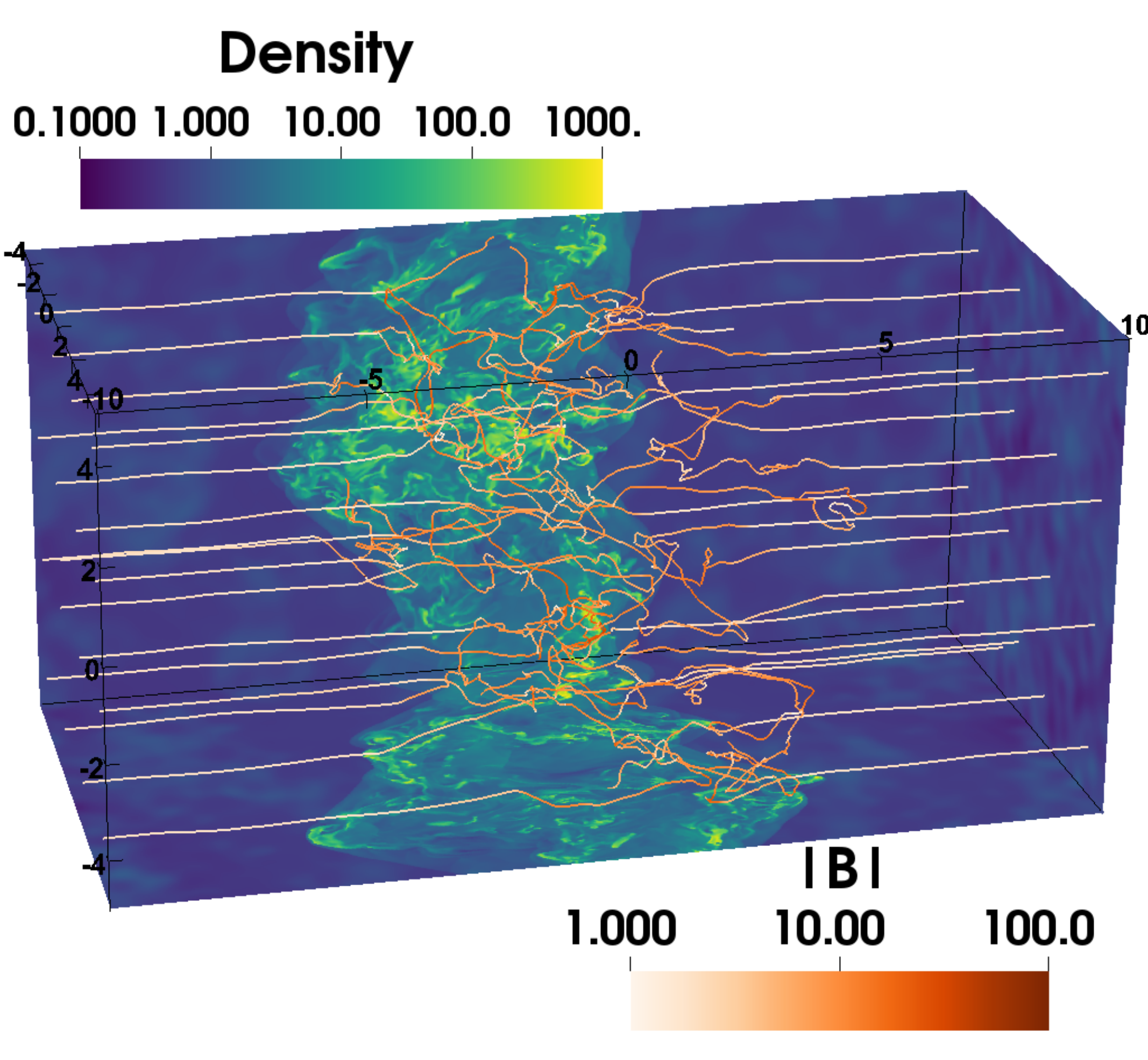}}
    \end{minipage}
    \begin{minipage}[t]{0.33\textwidth}
        \centering{\includegraphics[width = 1.00\linewidth]{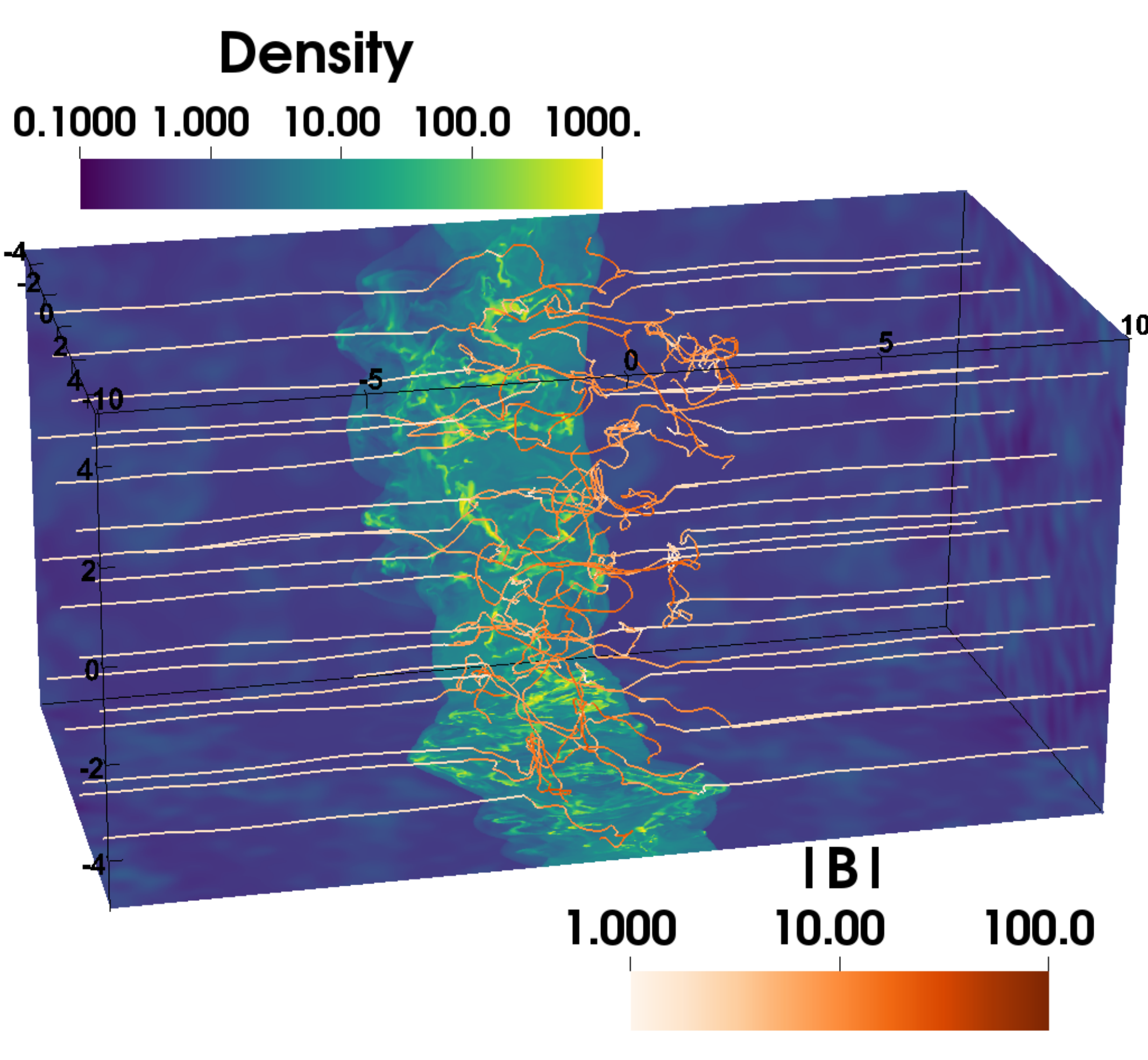}}
    \end{minipage}
    \begin{minipage}[t]{0.33\textwidth}
        \centering{\includegraphics[width = 1.00\linewidth]{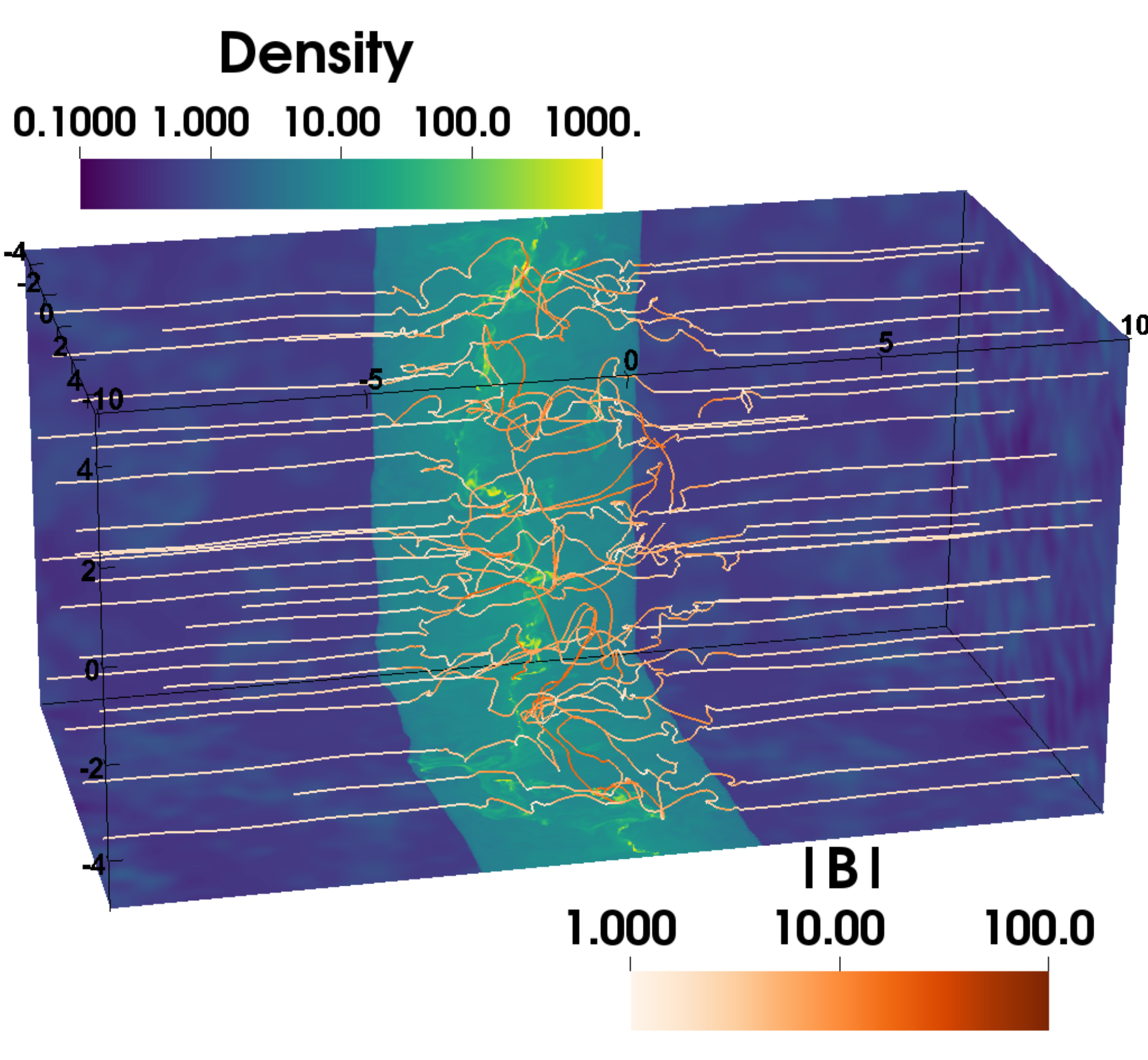}}
    \end{minipage}\\
    \begin{minipage}[t]{1.0\textwidth}
    \hfill\vspace{10pt}
    \end{minipage}\\
    \hspace*{6.2cm} {\large (d) 0.5 $\zsun$ at $t/t_{\rm cool}=3.0$}
    \hspace{1.2cm}  {\large (e) 0.2 $\zsun$ at $t/t_{\rm cool}=3.0$} \\
    \begin{minipage}[t]{0.33\textwidth}
        \hspace{1.3cm}
    \end{minipage}
    \begin{minipage}[t]{0.33\textwidth}
        \centering{\includegraphics[width = 1.00\linewidth]{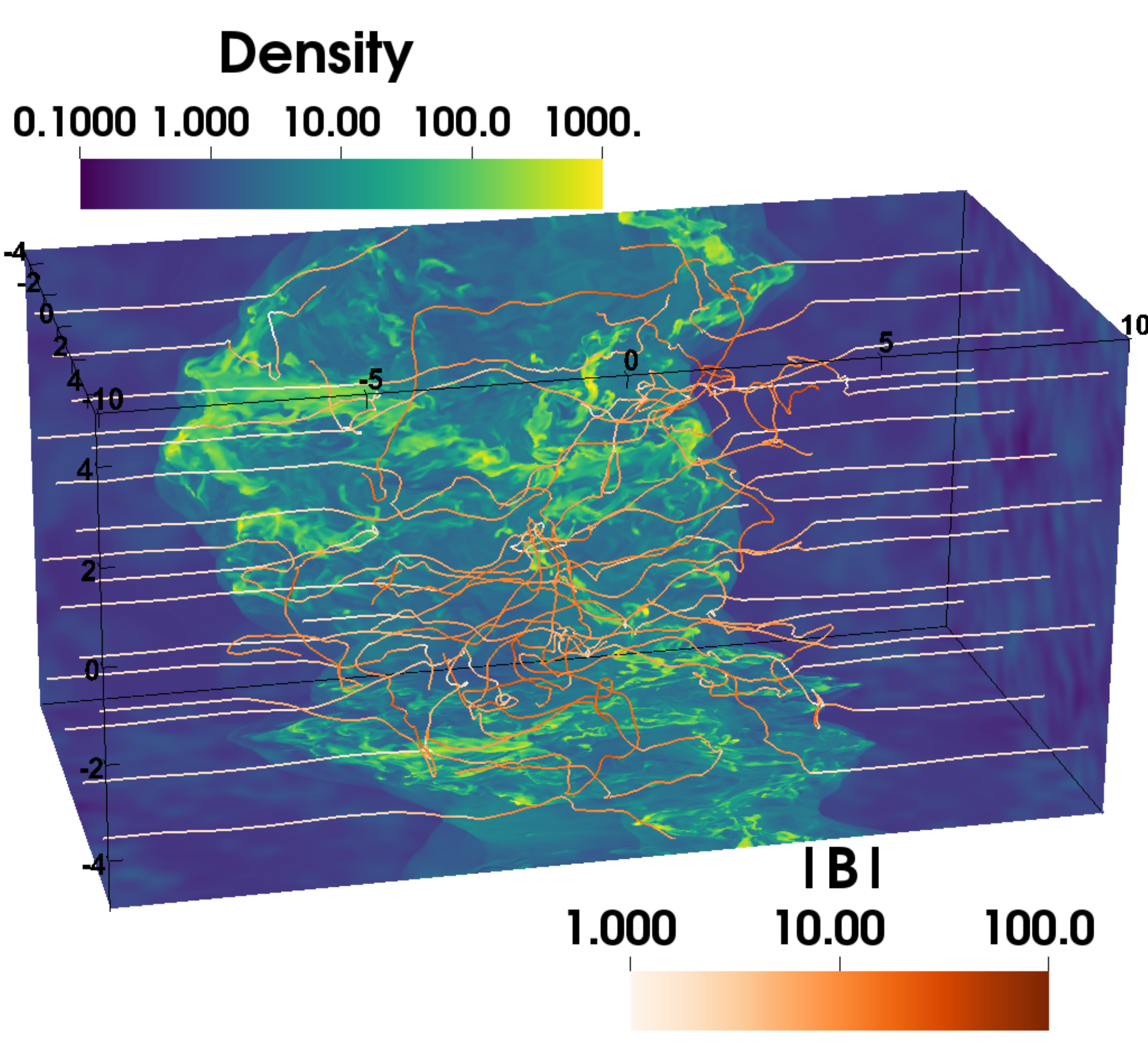}}
    \end{minipage}
    \begin{minipage}[t]{0.33\textwidth}
        \centering{\includegraphics[width = 1.00\linewidth]{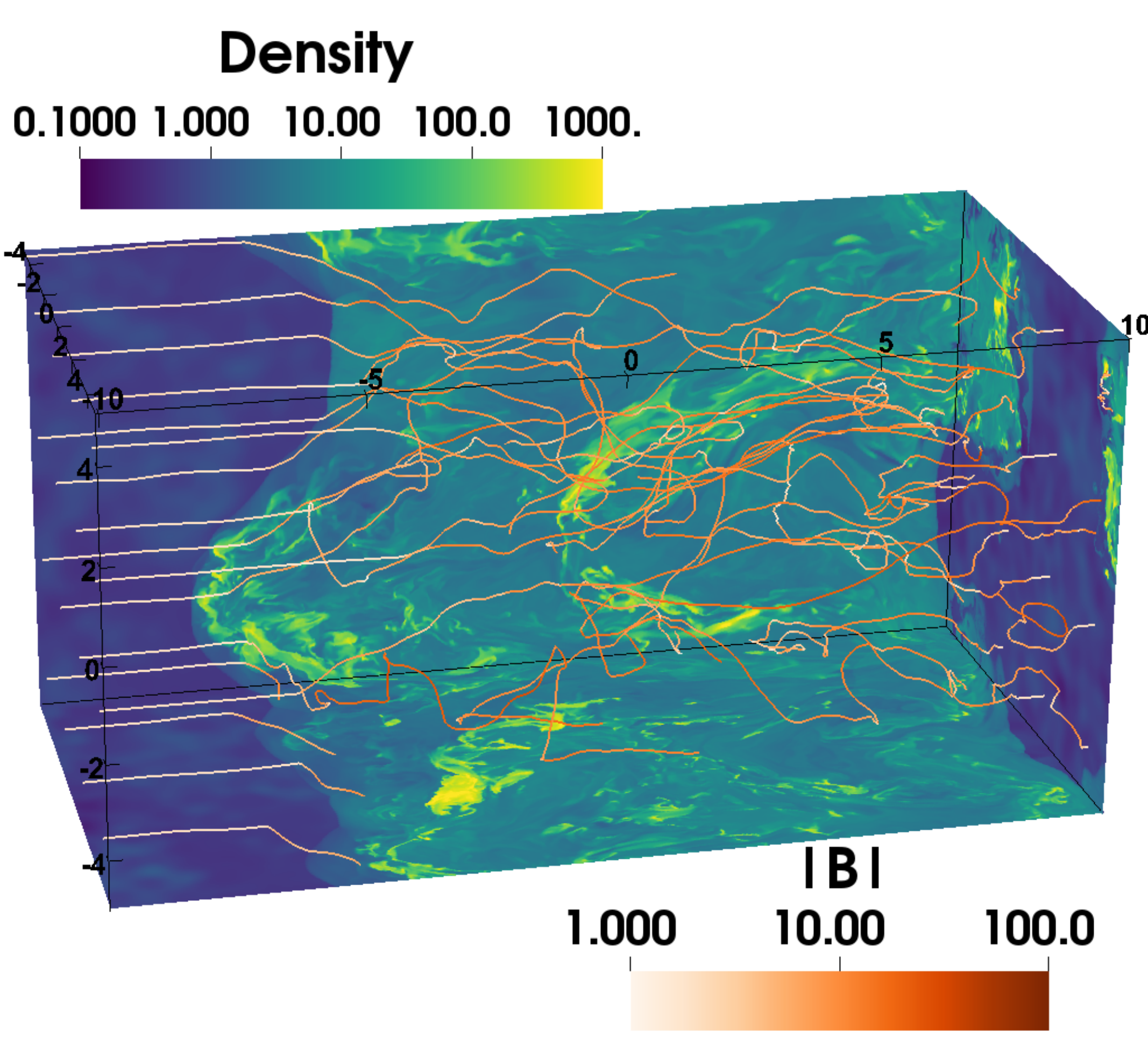}}
    \end{minipage}\\
    \caption{The three-dimensional view of the density and magnetic fields.
             Panels (a), (b), and (c) are at $t=3$ Myr from $1.0\, \zsun$, $0.5\, \zsun$, and $0.2\, \zsun$.
             Panels (d) and (e) are at $t = 3 t_{\rm cool}$ from $0.5\, \zsun$ and $0.2\, \zsun$,
             which therefore correspond to $t=6.0$ Myr and $15.0$ Myr, respectively.
             We show the slices of the density field on the three boundaries at $x=10$ pc, $y=-5$ pc, and $z=-5$ pc.
             The density is colored with $n\,[\mathrm{cm}^{-3}]$
             from blue ($0.1$ cm$^{-3}$) to yellow ($10^3$ cm$^{-3}$).
             The magnetic field lines are integrated from the two $x$ boundaries,
             whose strength $\left|B\right|$ is colored in orange.
             }
    \label{fig:DB3view}
\end{figure*}

\section{Results}
\label{sec:main}
\subsection{Expectations}
\label{subsec:expect}
Figure~\ref{fig:nPkeq} shows that
the CNM thermal state at $n>10^2\, \cmkk$ is similar 
within the $1.0$--$0.2\, \zsun$ range.
In this metallicity range, 
the cooling rate
is proportional to the metallicity \citep[see][]{Inoue2015}.
These suggest that the $t_{\rm cool}$ is longer 
in low-metallicity environments as $\propto Z^{-1}$
but CNM properties may become similar between 
$1.0$--$0.2\, \zsun$ at the same time measured in the unit of the cooling time.

The typical $t_{\rm cool}$ of the injected WNM is $\sim 1$ Myr at $1\, \zsun$,
$\sim 2$ Myr at $0.5\, \zsun$
and $\sim 5$ Myr at $0.2\, \zsun$ (See Section~\ref{subsec:tcool}).
\cite{Kobayashi2020} investigate the converging flow
at $1\, \zsun$ 
and suggest that the shock-compressed layer 
achieves a quasi-steady state at $\sim 3 t_{\rm cool}$.
Therefore, to make a comparison between different metallicities both at the same physical time and 
at the same time measured in the unit of the cooling time, 
we integrate until 3 $t_{\rm cool}$ in each metallicity.
Table~\ref{table:ParamsMet} lists these parameters.
\begin{figure*}
    \hspace{0.1cm} {\large (a) 1.0 $\zsun$ at $t/t_{\rm cool}=3.0$}
    \hspace{1.2cm} {\large (b) 0.5 $\zsun$ at $t/t_{\rm cool}=1.5$}
    \hspace{1.2cm} {\large (c) 0.2 $\zsun$ at $t/t_{\rm cool}=0.6$} \\
    \begin{minipage}[t]{0.33\textwidth}
        \centering{\includegraphics[width = 1.00\linewidth]{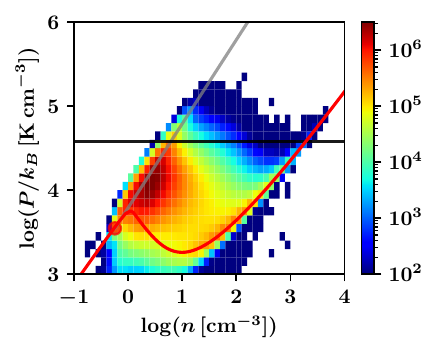}}
    \end{minipage}
    \begin{minipage}[t]{0.33\textwidth}
        \centering{\includegraphics[width = 1.00\linewidth]{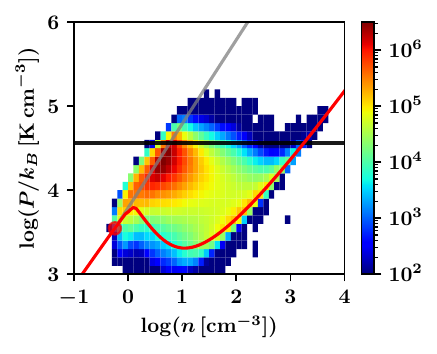}}
    \end{minipage}
    \begin{minipage}[t]{0.33\textwidth}
        \centering{\includegraphics[width = 1.00\linewidth]{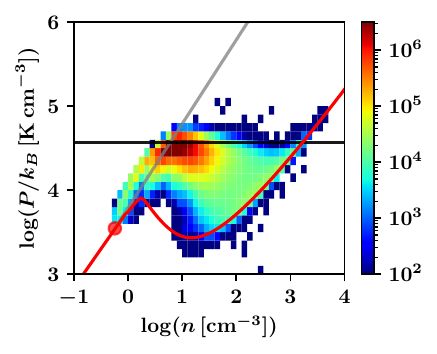}}
    \end{minipage}\\
    \begin{minipage}[t]{1.0\textwidth}
    \hfill\vspace{10pt}
    \end{minipage}\\
    \hspace*{6.2cm} {\large (d) 0.5 $\zsun$ at $t/t_{\rm cool}=3.0$}
    \hspace{1.2cm}  {\large (e) 0.2 $\zsun$ at $t/t_{\rm cool}=3.0$} \\
    \begin{minipage}[t]{0.33\textwidth}
        \hspace{1.3cm}
    \end{minipage}
    \begin{minipage}[t]{0.33\textwidth}
        \centering{\includegraphics[width = 1.00\linewidth]{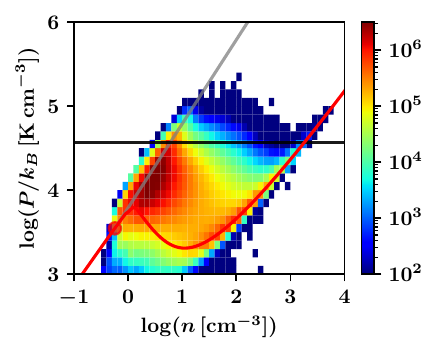}}
    \end{minipage}
    \begin{minipage}[t]{0.33\textwidth}
        \centering{\includegraphics[width = 1.00\linewidth]{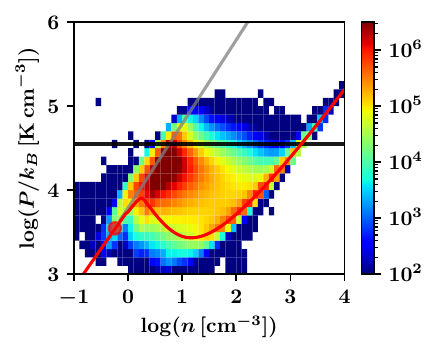}}
    \end{minipage}\\
    \caption{The phase diagram of the shock-compressed layer.
             The upper (lower) panels compare the thermal states at 
             $t=3$ Myr
             (at $t=3t_{\rm cool}$).
             The color represents the number of cells.
             The red circle at the lower left shows the initial state of the injected WNM
             and the gray solid line emanating from that red point
             shows the isothermal line (\ie, $P \propto n$).
             The black horizontal line shows the flow ram pressure.
    }
    \label{fig:TermStats}
\end{figure*}

\subsection{The thermal states and magnetic field evolution}
\label{subsec:thermal}
Figure~\ref{fig:DB3view} shows examples of the three-dimensional view of our simulation results.
Panels (a), (b), and (c) compare the results at 3 Myr from the $1.0\, \zsun$, $0.5\, \zsun$, and $0.2\, \zsun$ runs.
Panels (d) and (e) compare the results at $3t_{\rm cool}$ from the $0.5\, \zsun$ and $0.2\, \zsun$ runs.
These panels show that CNM clumpy/filamentary structures form slower in the lower metallicity environments.
The development of such CNM structures similar to that in $1\,\zsun$ environment
requires similar time measured in the units of the cooling time, for example,
at $3t_{\rm cool}$.
Compared at the same physical time of 3 Myr, the geometry of the shock-compressed layer 
in a lower metallicity environment 
is less disturbed (\eg, Panel (c)), close to a plane-parallel configuration 
because the inefficient cooling keeps the layer more adiabatic.

Figure~\ref{fig:TermStats} compares the thermal states of the different metallicities
at 3 Myr and at 3 $t_{\rm cool}$.
This shows that, at 3 Myr, 
the shock-compressed layer is still dominated more by the shock-heated WNM/UNM
in the lower metallicity environment.
Their thermal pressure is still close to the flow ram pressure.
Therefore, the temperature just starts to decrease 
in the shock-compressed layer of $0.2\,\zsun$ at 3 Myr (0.6 $t_{\rm cool}$),
so that the plane-parallel geometry of the shock-compressed layer
seen in Figure~\ref{fig:DB3view} 
is close to a simple one-dimensional shock compression
(see also Section~\ref{subsec:turbulent} and Appendix~\ref{sec:TV} for its impact on the turbulent velocity).

\begin{figure*}
    \hspace{0.1cm} {\large (a) 1.0 $\zsun$ at $t/t_{\rm cool}=3.0$}
    \hspace{1.2cm} {\large (b) 0.5 $\zsun$ at $t/t_{\rm cool}=3.0$}
    \hspace{1.2cm} {\large (c) 0.2 $\zsun$ at $t/t_{\rm cool}=3.0$} \\
    \begin{minipage}[t]{0.33\textwidth}
        \centering{\includegraphics[width = 1.00\linewidth]{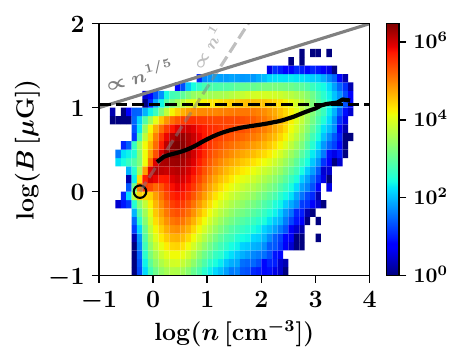}}
    \end{minipage}
    \begin{minipage}[t]{0.33\textwidth}
        \centering{\includegraphics[width = 1.00\linewidth]{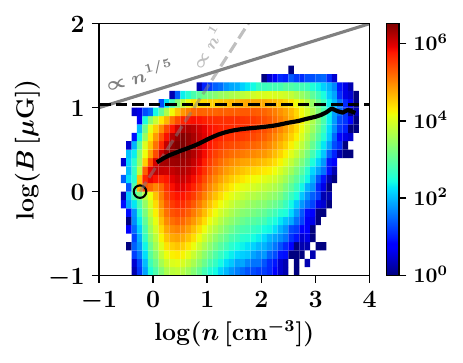}}
    \end{minipage}
    \begin{minipage}[t]{0.33\textwidth}
        \centering{\includegraphics[width = 1.00\linewidth]{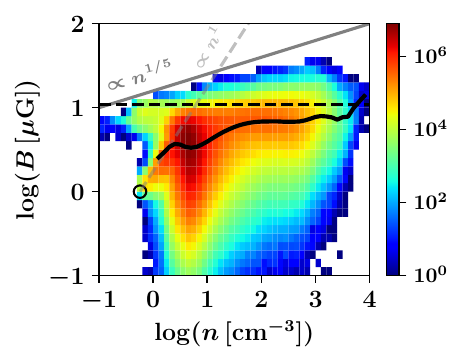}}
    \end{minipage}\\
    \begin{minipage}[t]{1.0\textwidth}
    \hfill\vspace{10pt}
    \end{minipage}\\
    \caption{The phase histogram on the plane of of the magnetic field strength and the number density. 
             The color represents the number of the numerical cells.
             The black small circle shows the initial condition of the injected WNM flow
             and the translucent dashed gray line emanating from this black circle shows
             the relation of $B\propto n^{1}$.
             The horizontal black dashed line shows the typical maximum field strength 
             by assuming that the balance between the magnetic pressure
             and the inflow ram pressure as Equation~\ref{eq:Beq}.
             The black solid curve shows the volume-weighted mean field strength $\langle B\rangle$ as a function of density
             in the range of $n\geq1\,\cmkk$.
             The gray straight line at the top shows the relation of $\langle B \rangle \propto n^{1/5}$.
    }
    \label{fig:Bn}
\end{figure*}

The net magnetic flux in our simulation does not increase in time because the mean magnetic field
is completely parallel to the WNM flow.
Nevertheless as we see in Figure~\ref{fig:DB3view},
the pre-shock density fluctuation induces a number of oblique shocks at the shock front,
which locally fold the field lines.
The magnetic fields in the shock-compressed layer 
are further twisted and stretched 
due to the turbulence,
which introduces a larger scatter in the local field strength.
We investigate the relation between the field strength and the number density at 3 $t_{\rm cool}$. 
Figure~\ref{fig:Bn} shows
the phase histogram on the plane of the magnetic field strength and the number density in the shock-compressed layer.
This figure shows that the field strength is initially amplified toward the shock-heated WNM volume through 
the shock compression almost following $B\propto n^{1}$ (as indicated with 
the translucent gray line).
The strength further varies due to the turbulence at $n< 10\,\cmkk$.
Note that the initial mean magnetic field is completely parallel to the flow velocity
and not all the volume experiences a perfect one-dimensional shock compression.
Therefore, the most frequent field strength is slightly weaker than that the relation
$B\propto n^{1}$ (see the upper panels of Figure~\ref{fig:Bn}).

Accompanying the density enhancement toward 
$n\sim 10^3\,\cmkk$ by the thermal instability,
the field strength gradually increases but with a limited level.
The black solid curve in Figure~\ref{fig:Bn} shows the volume-weighted average of the magnetic field strength 
as a function of the number density $\langle B \rangle (n)$,
where $B=\left| \mathbf{B} \right|$\footnote{Throughout this paper,
$\langle \cdot \rangle$ represents the volume-weighted average whereas $\langle\cdot\rangle_{\rho}$ represents the density-weighted average.}.
This shows that the amplification roughly scales with $\langle B \rangle \propto n^{1/5}$
in the $n>1\,\cmkk$ range.
This slow evolution occurs because condensing motion by the thermal instability 
is confined along the magnetic field orientations.
Similar results are 
previously obtained 
by previous numerical studies in the solar metallicity environment as well \citep[\eg,][]{Hennebelle2008,Heitsch2009,Inoue2012}.
Our results show that this gradual amplification
of the magnetic field occurs also in lower metallicity environments.

Figure~\ref{fig:Bn} indicates that there is a maximum magnetic field strength attained.
We can estimate this maximum strength 
by assuming the balance between the post-shock magnetic pressure with the ram pressure of the injected WNM as 
$B_{\rm eq}^2/8\pi = \rho_0 V_{\rm inflow}^2$.
This gives the typical value as 
\begin{eqnarray}
    B_{\rm eq} = 11 \, \mu{\rm G} \, \left( \frac{n_0}{0.57\,\cmkk} \right)^{1/2} \left( \frac{V_{\rm inflow}}{20\,\kms} \right)  \,.
    \label{eq:Beq}
\end{eqnarray}
Although this is an extreme case where the magnetic energy dominates the shock-compressed layer,
Equation~\ref{eq:Beq} should give a good approximation 
even when we consider local enhancement by the turbulence and the condensing motion by the thermal instability,
because the inflow ram pressure determines the typical maximum pressure of the shock-compressed layer.
$B_{\rm eq}$ shown in 
Figure~\ref{fig:Bn} (the horizontal black dashed line)
indeed outlines the typical maximum strength of the magnetic fields.

Note that the field strength can increase beyond $B_{\rm eq}$ at the densest volume where the self-gravity plays a role.
This occurs in $n\gtrsim 2.6\times 10^3\,\cmkk$ in our simulation setup.
Such a critical density of $n\sim 2.6\times 10^3\,\cmkk$
can be estimated as the density at which
the CNM plasma beta with $B = B_{\rm eq}$
becomes the unity \citep{Iwasaki2022}.
It is difficult to clearly confirm this amplification in our current simulations due to our limited volume 
and due to our diffuse WNM-only initial condition.
Nevertheless, we expect that the field strength increases also in the low metallicity environment 
because the CNM thermal states are similar between 1.0--0.2 $\zsun$
so are the critical densities at which the self-gravity starts to dominate.
Previous simulations of a converging flow at $1.0\,\zsun$ show $\langle B \rangle \propto n^{1/2}$ at $n>10^3\,\cmkk$
either when they integrated much longer time to accumulate mass or when they started with two-phase atomic flows
with their mean number density already $n\gtrsim 5\,\cmkk$ \citep[\eg,][]{Hennebelle2008,Inoue2012,Iwasaki2022}.

In conclusion of this Section~\ref{subsec:thermal},
our results show that
the development of CNM clumpy/filamentary structure 
and the magnetic field strength is
similar between metallicities at the same time measured in the unit of the cooling time 
(\ie, the evolution slows down linearly with the metallicity in terms of the physical time).
The field strength is relatively constant with a few $\mu$G to 10 $\mu$G 
up to $n<10^3\,\cmkk$ as $\langle B \rangle \propto n^{1/5}$,
and possibly starts to amplify further by the self-gravity.
Note that this field strength in $n<10^3\,\cmkk$ is consistent with Zeeman measurements of low (column) density
regions of molecular clouds in the Milky Way
\citep[\eg,][]{Crutcher2012}
and Faraday rotation measurements toward the ISM in the LMC and SMC \citep{Gaensler2005,Rainer2013,Livingston2022}.
\begin{figure*}
    \hspace{0.3cm} {\large (a) Velocity Dispersion }
    \hspace{2.2cm} {\large (b) Mean Density        }
    \hspace{2.2cm} {\large (c) Turbulent Pressure  } \\
    \begin{minipage}[t]{0.37\textwidth}
        \centering{\includegraphics[width = 1.0\linewidth]{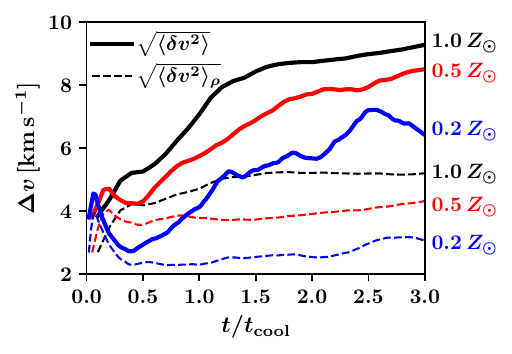}}
    \end{minipage}
    \begin{minipage}[t]{0.31\textwidth}
        \centering{\includegraphics[width = 1.0\linewidth]{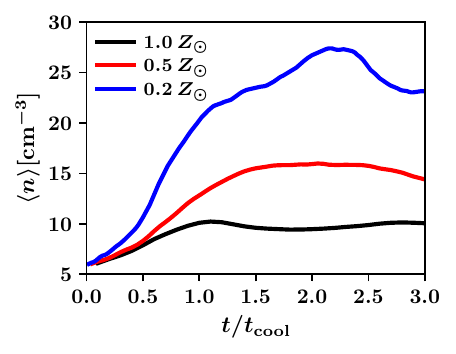}}
    \end{minipage}
    \begin{minipage}[t]{0.31\textwidth}
        \centering{\includegraphics[width = 1.0\linewidth]{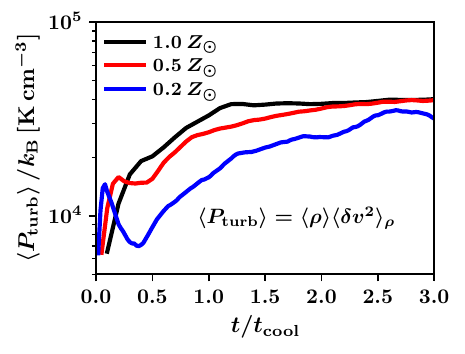}}
    \end{minipage}\\
    \caption{The $t_{\rm cool}$-normalized time evolution of the volume-weighted velocity dispersion $\sqrt{\langle\delta v^2\rangle}$
             and the density-weighted velocity dispersion $\sqrt{\langle\delta v^2\rangle_{\rho}}$ (Panel (a)), 
             the volume-weighted mean density $\langle n \rangle$ (Panel (b)),
             and the averaged turbulent pressure $\langle P_{\rm turb} \rangle$ (Panel (c)),
             where $\langle P_{\rm turb} \rangle = \langle \rho \rangle \langle \delta v^2 \rangle_{\rho}$.
             The black, red, and blue colors represent $1.0$, $0.5$, and $0.2\,\zsun$, respectively.
    }
    \label{fig:Turbulence}
\end{figure*}

\subsection{The overall turbulent structure}
\label{subsec:turbulent}
In \cite{Kobayashi2022}, we show that 
the coevolution of the turbulence and the thermal evolution by the thermal instability plays a significant role to 
determine the multiphase density structure and its density probability distribution function at the molecular cloud formation stage.
\cite{Kobayashi2022}, however, neglect the magnetic fields 
and studied simple hydrodynamic converging flows.
In this section, we will confirm those findings even in the current magnetohydrodynamics simulations 
and investigate how the turbulent structure differs/resembles between the cases with different metallicities.

Panels (a), (b), and (c) 
of Figure~\ref{fig:Turbulence} show the evolution of the velocity dispersion, the mean density, and the turbulent pressure,
respectively, as a function of the time measured in the unit of the cooling time (\ie, $t/t_{\rm cool}$).
In the early stage at $t<0.5\, t_{\rm cool}$,
the shock-compressed layer has plane-parallel geometry in the low metallicity environment
(see Section~\ref{subsec:thermal}).
Due to the resultant efficient deceleration of the inflow and the 
inefficient cooling, 
the mass is dominated by the shock-heated WNM state with slow turbulence,
with negligible volume/mass of the CNM\@.
Therefore, the volume-weighted velocity dispersion $\sqrt{\langle\delta v^2\rangle}$
is close to the density-weighted velocity dispersion $\sqrt{\langle\delta v^2\rangle_{\rho}}$
of $\sim 2\,\kms$ in $0.2\,\zsun$ (Panel (a) of Figure~\ref{fig:Turbulence}; 
see also Appendix~\ref{sec:TV}).

At $t\gtrsim 0.5\, t_{\rm cool}$
in the lower metallicity environment, 
the phase transition from the shock-heated WNM to the CNM 
tends to occur at
a pressure closer to the inflow ram pressure
(see Panel (c) Figure~\ref{fig:TermStats}).
The mean number density is
higher in lower metallicity environments accordingly (Panel (b) of Figure~\ref{fig:Turbulence}).
During this evolution, the turbulent pressure gradually grows until it balances against the inflow ram pressure.
As a result, the turbulent pressure is almost the same between the three metallicities at 3 $t_{\rm cool}$ 
(Panel (c) of Figure~\ref{fig:Turbulence}).

These results suggest that 
the turbulent pressure is almost the same at 3 $t_{\rm cool}$.
The difference exists due to the efficient compression in the lower metallicity environments, 
because the inflow and the shock front incidents with almost 90 deg angle,
which results in the denser postshock WNM with slower velocity.
However, this difference is limited to by a factor of 2 (4) in the velocity dispersion (in the mean density, respectively).
The value of $\sqrt{\langle\delta v^2\rangle} \sim 6$--$9\,\kms$, close to the sound speed of the WNM,
suggests that the turbulence on the molecular cloud scale is powered by the WNM super-Alfv\'{e}nic turbulence 
not only in the Milky Way galaxy but also in the LMC and SMC\footnote{Note that the turbulence at $n\sim 10^3\,\cmkk$ is typically super-Alfv\'{e}nic.
$\sqrt{\langle\delta v^2\rangle}$ ($\sqrt{\langle\delta v^2\rangle_{\rho}}$)
roughly traces the turbulence of the WNM (CNM) because the volume is dominated by 
the WNM whereas 50 percent of the mass is locked in the CNM \citep{Kobayashi2020}.
Therefore, combined with the slow evolution of $\langle B \rangle$ with $n$,
the turbulent Alfv\'{e}nic Mach number typically ranges $\mathcal{M}_{A} \simeq 1.6$--$9.9$.}.

\begin{figure*}
    \hspace{0.1cm} {\large (a) 1.0 $\zsun$ at $t/t_{\rm cool}=3.0$}
    \hspace{1.2cm} {\large (b) 0.5 $\zsun$ at $t/t_{\rm cool}=3.0$}
    \hspace{1.2cm} {\large (c) 0.2 $\zsun$ at $t/t_{\rm cool}=3.0$} \\
    \begin{minipage}[t]{0.33\textwidth}
        \centering{\includegraphics[width = 1.00\linewidth]{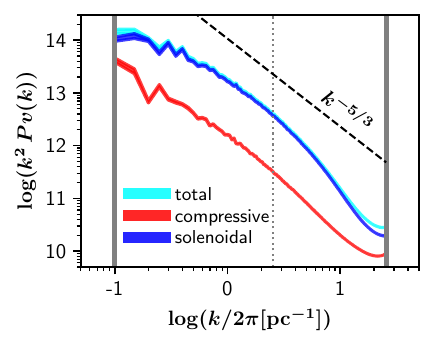}}
    \end{minipage}
    \begin{minipage}[t]{0.33\textwidth}
        \centering{\includegraphics[width = 1.00\linewidth]{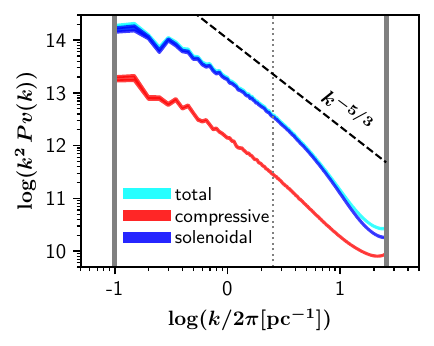}}
    \end{minipage}
    \begin{minipage}[t]{0.33\textwidth}
        \centering{\includegraphics[width = 1.00\linewidth]{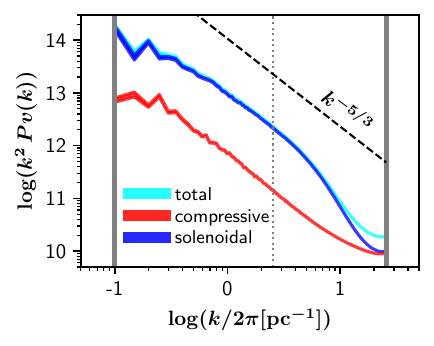}}
    \end{minipage}\\
    \caption{The one-dimensional averaged turbulent power spectrum of the shock-compressed layer
             at $t = 3 t_{\rm cool}$.
             The cyan, blue, and red curves correspond to the total power,
             the solenoidal mode power, and the compressive mode power, respectively.
             Their width represents the Poisson noise at each frequency $k$.
             It is difficult to visually recognize the cyan curve because the solenoidal mode dominates the total power so that
             they almost overlap in this presentation.
             The two vertical gray lines show the lowest frequency of the shock-compressed layer
             and the Nyquist frequency.
             The thin gray dotted line shows the one-tenth of the Nyquist frequency,
             above which the numerical diffusion impacts the Fourier power.
             We employ the powers between the left gray line ($k/2\pi=0.1$ pc$^{-1}$) 
             and the thin dotted line ($k/2\pi=2.56$ pc$^{-1}$)
             to measure the total fraction of the solenoidal and compressive modes
             in the subsequent analyses and figures.
             The black dashed line shows the one-dimensional Kolmogorov spectrum of $k^2 P_{v}(k) \propto k^{-5/3}$.
    }
    \label{fig:Pkv}
\end{figure*}
\begin{figure}
    \centering{\includegraphics[width = 1.00\linewidth]{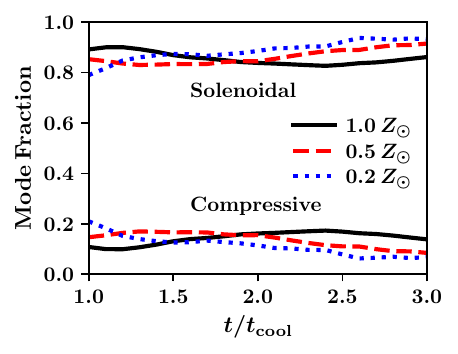}}
    \caption{The $t_{\rm cool}$-normalized time evolution of the turbulence mode fraction.
             The upper (lower) three lines correspond to the solenoidal (compressive, respectively) mode.}
    \label{fig:PkTev}
\end{figure}

To understand the turbulence structure, we perform the Fourier analysis of the turbulence by using the Fast Fourier Transform in the West
(FFTW 3.3; \citealt{FFTW2005}),
and decompose the turbulence into the solenoidal and compressive modes,
which 
are defined respectively as 
\begin{eqnarray}
    \mathbf{\tilde{v}}_{\rm sol}(\unitk)  &=&  \left( \unitk \times \mathbf{\tilde{v}} \right)  \times \unitk \,, \\
    \mathbf{\tilde{v}}_{\rm comp}(\unitk) &=&  \left( \unitk \cdot  \mathbf{\tilde{v}} \right)         \unitk \,.
\end{eqnarray}
Here, $\unitk$ represents a unit wave vector and $\mathbf{\tilde{v}}$ represents the Fourier component of the velocity field.
We here denote $k=\left|\mathbf{k}\right|$.
Figure~\ref{fig:Pkv} shows the power spectrum
at $3\,t_{\rm cool}$.
The solenoidal mode dominates the turbulence power on all scales.
Table~\ref{table:ModeFrac} summarizes the total fraction of each turbulent mode at $3\,t_{\rm cool}$.
The solenoidal mode fraction amounts to 80--90 percent.
In calculating this, 
we employ the powers between $k/2\pi=0.1$ and $2.56$ pc$^{-1}$ 
to avoid numerical diffusion effects on small scale
(we refer the readers to the caption of Figure~\ref{fig:Pkv})\footnote{We hereafter 
denote the spatial frequency $k/2\pi$ as the inverse of the wavelength
so that the wave with $k/2\pi=0.1$ pc$^{-1}$ has the wavelength of 10 pc.}.
\begin{table}
    \caption{Mode fraction at $3\,t_{\rm cool}$, where the total power is normalized to the unity.}
    \centering{
        \begin{tabular}{c||c|c|c}
            \hline
            \hline
            Metallicity  & $1.0\,\zsun$     & $0.5\,\zsun$  & $0.2\,\zsun$ \\ \hline 
            Solenoidal   & 0.86                & 0.91             & 0.93  \\
            Compressive  & 0.14                & 0.09             & 0.07   \\
            \hline
            \hline
        \end{tabular}
    }    
    \label{table:ModeFrac}
\end{table}
Figure~\ref{fig:PkTev} shows the time evolution of this mode fraction
as a function of $t/t_{\rm cool}$.
The solenoidal (compressive) mode fraction evolves quasi-steadily 
with $\sim 80$ percent (20 percent, respectively) after $1\,t_{\rm cool}$.

Such solenoidal-mode-dominated turbulence is a natural consequence of
the accreting system followed with the thermal instability,
as we discuss in our previous converging flow simulations at $1.0\,\zsun$ \citep{Kobayashi2022}.
At early stages, 
the shock fronts deform by the interaction with the inflow density inhomogeneity.
This curved shock fronts introduce inhomogeneity in the postshock entropy distribution.
Such entropy inhomogeneity generates the turbulent vorticities accounting for
a small fraction of the solenoidal mode \citep[\cf,][]{Kida1990a}.
At later stages after $1.0\,t_{\rm cool}$,
the interaction between the formed CNM clumps and the shock fronts
significantly deforms the shock fronts.
This deformation creates a number of oblique shocks, whose
maximum size is comparable to the size of the shock-compressed layer.
This induces strong shear motion into the shock-compressed layer,
so that the solenoidal mode dominates the turbulence.
Our results show that the solenoidal-mode dominated turbulence emerges
also in the low metallicity environment. 

In conclusion of this Section~\ref{subsec:turbulent},
our results suggest that
in the range of $1.0$--$0.2\,\zsun$,
molecular clouds forming in the shock-compressed layer 
have the turbulent pressure close to the inflow ram pressure.
$\gtrsim 80$ percent of the turbulence power is
in the solenoidal mode at all the metallicities.
This indicates that 
even in a galactic-scale converging region,
forming molecular clouds are always solenoidal-mode dominated.
Therefore, a galactic-scale compressive motion
is important to form molecular clouds 
but it does not 
immediately  
mean enhancement of star formation efficiency by enhancing compressive motion in molecular clouds.

\subsection{The properties/statistics of the CNM clumps}
\label{subsec:CNM}
We identify the CNM structures to further investigate their properties.
In this section, we define the CNM as the volume with 
$T<200$ K and $n>20\,\cmkk$.

\begin{figure*}
    \begin{minipage}[t]{0.50\textwidth}
        \centering{\includegraphics[width = 1.00\linewidth]{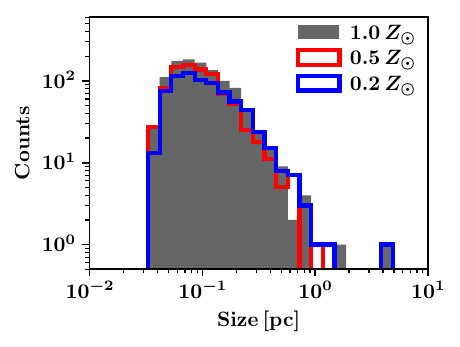}}
    \end{minipage}
    \begin{minipage}[t]{0.50\textwidth}
        \centering{\includegraphics[width = 1.00\linewidth]{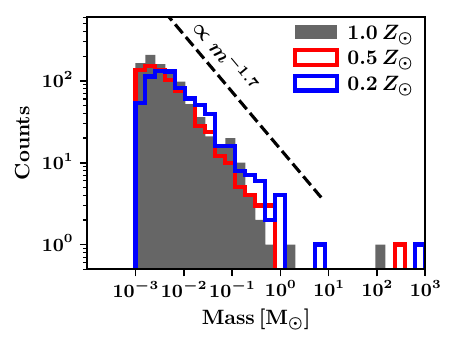}}
    \end{minipage}\\
    \caption{The size and mass functions of CNM clumps. The grey filled histogram shows the $1.0\,\zsun$
             and the other step histograms show the $0.5\,\zsun$ (red) and $0.2\,\zsun$ (blue).
             The mass function shows a reference power-law distribution of ${\rm d}n/{\rm d}m \propto m^{-1.7}$ (black dashed line).
             }
    \label{fig:CNMMF}
\end{figure*}

First, we perform the Friends-of-Friends algorithm 
to identify CNM clumps
as groups of connected CNM cells.
Each clump has more than 64 member cells
to avoid numerical noise on small scales.
Figure~\ref{fig:CNMMF} compares the size and mass functions of the identified CNM clumps
at 3 $t_{\rm cool}$.
We define the size $l$ of each CNM clump as $l=\sqrt{I_{\rm max}/M}$
where $I_{\rm max}$ and $M$ are the maximum eigenvalue
of its inertia matrix and the mass of each clump.
The size distribution peaks at $\sim 0.1$ pc at all the metallicities\footnote{The thermal instability grows 
also on scales smaller than 0.1 pc, but the sharp cutoff on $< 0.1$ pc is originated from 
our criteria of $\geq$ 64 member cells to avoid any numerical noise on that scale.}.
The mass functions follow the power-law distribution of ${\rm d}n/{\rm d}m \propto m^{-1.7}$
up to $\sim 10^2\,\msun$.
Some CNM mass tends to stagnate in the central region of the shock-compressed layer due to the converging flow configuration (see Figure~\ref{fig:DB3view} and Appendix~\ref{sec:TV}).
Large CNM clumps obtain more mass or coagulate with other CNM gas so that the most massive clumps deviate from the power-law distribution.
\begin{figure*}
    \hspace{0.3cm} {\large (a) 1.0 $\zsun$ at 3  Myr} 
    \hspace{2.3cm} {\large (b) 0.5 $\zsun$ at 6  Myr} 
    \hspace{2.3cm} {\large (c) 0.2 $\zsun$ at 15 Myr}  \\
    \begin{minipage}[t]{0.33\textwidth}
        \centering{\includegraphics[width = 1.00\linewidth]{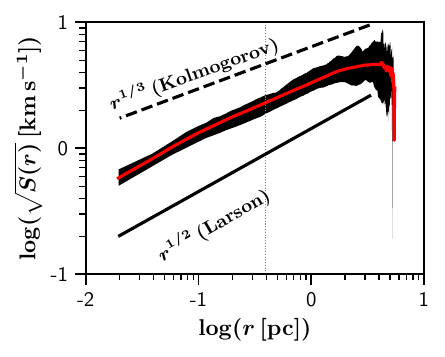}}
    \end{minipage}
    \begin{minipage}[t]{0.33\textwidth}
        \centering{\includegraphics[width = 1.00\linewidth]{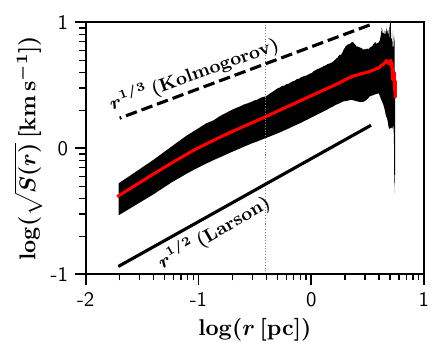}}
    \end{minipage}
    \begin{minipage}[t]{0.33\textwidth}
        \centering{\includegraphics[width = 1.00\linewidth]{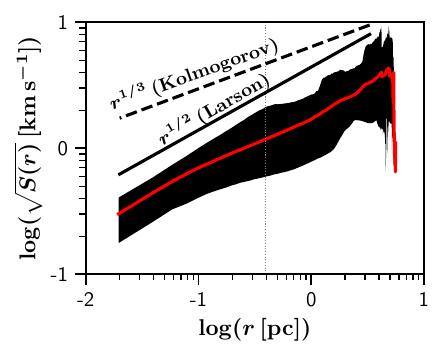}}
    \end{minipage}\\
    \caption{The square root of the second-order velocity structure function $S(r)$ as a function of the distance $r$ between CNM cells
             at 3 $t_{\rm cool}$.
             The black shaded region shows the maximum and minimum of $S(r)$ among the sampled 27 sub-volumes.
             The red solid curve shows the volume-weighted average of all the sub-volumes.
             The black solid (dashed) line shows the Larson (Kolomogorov) power-law relation as $\sqrt{S(r)} \propto r^{1/2}$ 
             ($\sqrt{S(r)} \propto r^{1/3}$).
             The vertical grey dotted line at $k/2\pi=0.4$ pc$^{-1}$ indicates the one tenth of the Nyquist frequency
             (\ie, the typical spatial scale below which the turbulent power is affected by the numerical diffusion).}
    \label{fig:vsf}
\end{figure*}
The index $-1.7$ has often been reported in the $1.0\,\zsun$ environment
by previous converging WNM flow simulations 
(in two-dimension \citep[\eg][]{Heitsch2005,Hennebelle2007a,Hennebelle2008}
and in three-dimension \citep[\eg][]{Inoue2012}).
This power-law function is explained from the statistical growth of the thermal instability
with a given density fluctuation spectrum. 
The functional form of ${\rm d}n/{\rm d}m \propto m^{(\alpha-3)/3-2}$ is expected 
when the seed density fluctuation has a three-dimensional averaged spectral index of $\alpha$
\citep[we refer the readers to see][for the detailed explanation 
and the derivation of the index $(\alpha-3)/3-2$]{Hennebelle2007a}.
When $\alpha=11/3$, \ie, the Kolmogorov fluctuation, 
this gives ${\rm d}n/{\rm d}m \propto m^{-1.7}$, 
which is indeed consistent with 
the numerical results of the previous authors in $1.0\,\zsun$ environments.
Our results suggest that,
at the same $t/t_{\rm cool}$,
the same
CNM mass spectrum can appear also in the lower metallicity environments due to the thermal instability
with the Kolmogorov turbulence background. 

Second, we investigate the turbulence structure of the CNM volume alone by measuring
the two point correlation of the CNM velocity field.
We measure the second-order velocity structure function 
\begin{equation}
    S(r) = \langle \left| v(\mathbf{r}+\mathbf{x}) - v(\mathbf{x}) \right|^2 \rangle \,,
    \label{eq:vsf}
\end{equation}
where $r=\left|\mathbf{r}\right|$ and $x$ is the three-dimensional position of the CNM cells.
We select 27 sub-volumes in the shock-compressed layer, where 
each sub-volume is a $(3.3\, \mathrm{pc})^3$ cube 
and their volume-centered position is 
$(x,y,z)=(0\pm3.3\,\mathrm{pc},
          0\pm3.3\,\mathrm{pc},
          0\pm3.3\,\mathrm{pc})$.
Figure~\ref{fig:vsf} shows $\sqrt{S(r)}$ from each metallicity at 3 $t_{\rm cool}$.
The CNM volume overall shows the transition towards the Larson-type scale-dependence as 
$\sqrt{S(r)}\propto r^{1/2}$.
This relation extends from small scales within individual CNM clumps to large scales beyond those clumps.
This indicates that the commonly observed Larson's law in molecular clouds is
inherited from the CNM phase and is ubiquitous also in low metallicity environment down to $0.2\,\zsun$.

Note that the dynamic range is still limited if we consider only the scales
without the numerical diffusion (\ie, the scales larger than the vertical dotted line).
$S(r)$ in this range is between Kolmogorov and Larson's relations. Further high-resolution simulations
are required to finally conclude that CNM velocity structure function follows the Larson-type relation on all scales.
Nevertheless, 
the strongest turbulence power within the CNM clumps
is in eddies whose scale is comparable to the typical CNM clump size $\sim 0.1$ pc.
Our results, therefore, suggest that the internal velocity dispersion within the CNM clumps
remains $\lesssim 1$ km s$^{-1}$
while the clump-to-clump relative velocity
is $3$ -- $5$ km s$^{-1}$. 

Note that 
the amplitude of $\sqrt{S(r)}$ at $0.2\,\zsun$ is smaller by a factor of two than that at $1.0\,\zsun$,
as we see in $\sqrt{\langle \delta v^2 \rangle_{\rho}}$
(see Figure~\ref{fig:Turbulence}).
This is consistent with the comparison of the CO line width 
between clouds in the LMC/SMC and the Milky Way \citep{Bolatto2008,Fukui2010,Ohno2023},
except for those in extreme star-forming systems, such as 30 Dor R136 cluster \citep{Indebetouw2013}.
Also note that, a variation at a given spatial displacement $r$ is larger in the lower metallicity environment
because the shock-compressed layer is thicker and thus
the total number of cells in this analysis increases toward lower metallicity.

In conclusion of this Section~\ref{subsec:CNM},
our results show that 
CNM mass spectrum is ${\rm d}n/{\rm d}m \propto m^{-1.7}$ 
commonly in the $1.0$--$0.2\,\zsun$ range
with the Kolmogorov turbulence background. 
The second-order structure function 
of the CNM volume alone indicates the transition towards the Larson-type turbulence scale-dependence ubiquitously at all metallicities
with its amplitude smaller by a factor of two at $0.2\,\zsun$ compared with $1.0\,\zsun$.
\linebreak

\begin{figure}
    \centering{\includegraphics[width = 1.00\linewidth]{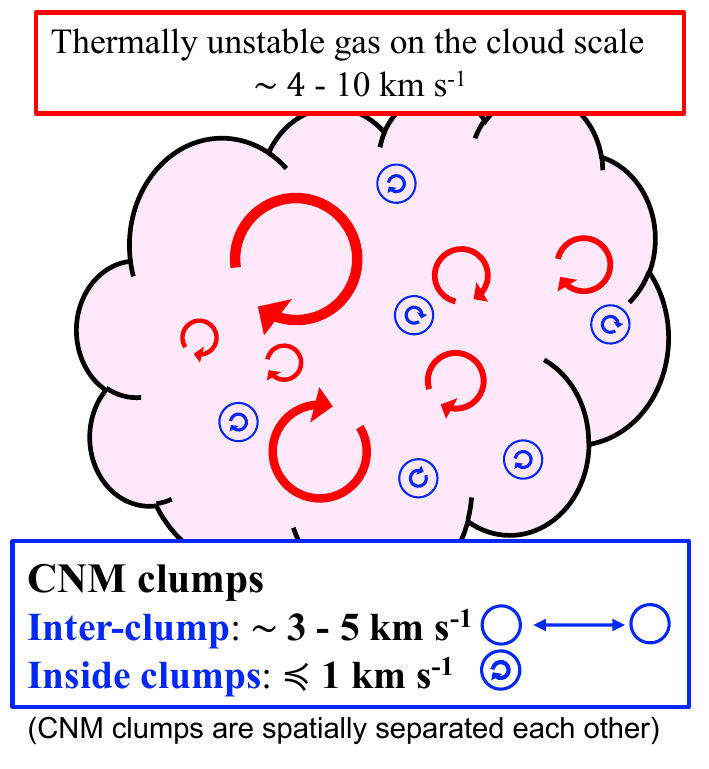}}
    \caption{A schematic figure showing the relationship between the density structures and their thermal states
             in the metallicity range of $1.0$--$0.2\,\zsun$.
             The WNM/UNM turbulence on the entire cloud scale powers the relative velocity between CNM clumps
             with $3$ -- $5$ km s$^{-1}$, while the internal velocity dispersion within individual CNM clumps
             remains $\lesssim 1$ km s$^{-1}$. 
             The velocity two-point correlation of the CNM volume follows the Larson's law 
             as $\sqrt{S(r)}\propto r^{1/2}$.}
    \label{fig:sch_MC}
\end{figure}

Combining all the results in Section~\ref{sec:main},
Figure~\ref{fig:sch_MC} schematically summarizes the hierarchical thermal/turbulent structure 
of the multi-phase medium in this metallicity range.

\section{Implications and Discussions}
\label{sec:imp}
Our results show that the physical properties of a shock-compressed layer 
scales with the metallicity in the $1.0$--$0.2\,\zsun$ range.
This suggests that the properties of subsequently-formed molecular clouds are 
likely similar between the Milky Way, LMC, and SMC if we somehow select and compare clouds
at the same $t/t_{\rm cool}$.

\subsection{Pre-existence of CNM structures in the WNM: CNM determines the formation of molecular clouds?}
\label{sec:preCNM}
On one hand in the $1.0\,\zsun$ context, previous authors often investigate the cloud formation by supersonic flows
in supernova remnants, galactic spirals \etc \citep{KimWT2020,Chevance2022}.
For example, a single supernova remnant expands to 50 pc size in 0.3 Myr ($\sim 0.3\,t_{\rm cool}$) 
and accumulates $10^4\,\msun$ H{\sc i} mass \citep[\eg,][]{Kim2015}.
The compilation of multiple supernovae events is able to create even more massive molecular clouds \citep{Inutsuka2015}.
On the other hand, in a low metallicity environment, 
the molecular cloud formation out of the WNM alone requires longer physical time.
This is the case even in the configuration where 
the mean magnetic field is completely parallel to the WNM inflow as 
we have studied,
which forms molecular clouds most efficiently compared with the magnetic fields inclined against the flow \citep{Inoue2008,Iwasaki2019}.

To estimate how this timescale has an impact on molecular cloud formation in a low metallicity environment, 
let us assume that all the CNM volume eventually evolves to the molecular gas.
Based on our simulation results\footnote{The CNM mass 
in the shock-compressed layer at 15 Myr at $0.2\,\zsun$ is 
$1.1\times10^3\,\msun$ in our simulation.},
the expected mass of the cloud, $M_{\rm cloud}$, is approximately
\begin{widetext}
\begin{align}
    &M_{\rm cloud} 
    \simeq 1.1\times 10^3\,\msun  \nonumber \\
    &\hspace{1em} \left(\frac{n_0}{0.57\,\cmkk} \right)
                   \left(\frac{V_{\rm inflow}}{20\,\kms}\right) 
                   \left(\frac{L}{10\,\mathrm{pc}} \right)^2 
                   \left(\frac{t}{3 t_{\rm cool}} \right) \label{eq:mcloud1} \\
    &\propto 
                   \left(\frac{n_0}{0.57\,\cmkk} \right)^2
                   \left(\frac{V_{\rm inflow}}{20\,\kms}\right) 
                   \left(\frac{L}{10\,\mathrm{pc}} \right)^2 
                   \left(\frac{Z}{0.2\zsun} \right)
                   \left(\frac{t}{15\,\mathrm{Myr}} \right) \label{eq:mcloud2} \,.
\end{align}
\end{widetext}
Here $L^2$ is the inflow cross section in our calculation domain
and thus 
the first three factors in Equation~\ref{eq:mcloud1} come from the inflow mass flux.
The exact dependence of $t_{\rm cool}$ on $n_0$ and $V_{\rm inflow}$ 
is still left for future studies,
which is involved in the conversion from Equation~\ref{eq:mcloud1} to Equation~\ref{eq:mcloud2}.
We here employ $t_{\rm cool} \propto n_0^{-1} Z^{-1}$
as the fiducial dependence, which is consistent with our definition of $t_{\rm cool}$ as calculated in Table~\ref{table:ParamsMet}
(see Section~\ref{subsec:deptcool} and Equation~\ref{eq:dep1}).

If we suppose the case that 
a single flow event with $n_0=0.57\,\cmkk$ and $V_{\rm inflow}=20\,\kms$ creates 
a typical maximum cloud of a few $10^5 \msun$ in the LMC/SMC,
Equation~\ref{eq:mcloud2} indicates that such a WNM flow should continue coherently on scales of $L\simeq 100$ pc and 15 Myr.
15 Myr is one order of magnitude longer compared with the typical expansion timescale of a single supernova remnant.
A superbubble rather than a single supernova event is more likely to keep such a coherent one-directional flow
over 15 Myr timescale.
However, prevalent existence of molecular clouds that is not associated with superbubbles 
in the LMC/SMC 
indicates that faster flows and/or pre-existence of CNM structure in the WNM
is important
for the cloud formation and the evolution in a lower metallicity environment \citep[\cf,][]{Dawson2013}\footnote{Note that 
the CNM formation efficiency is higher in the lower metallicity environments as we discussed in Section~\ref{subsec:thermal}.
This variation is, however, limited compared with the difference of the cooling time between metallicities
(\ie, $Z^{-1}$ in Equation~\ref{eq:mcloud1}).
In our simulations, the total mass of the CNM at 3$t_{\rm cool}$ are
$69\,\msun$ at $1\,\zsun$, 
$174\,\msun$ at $0.5\,\zsun$,
and $694\,\msun$ at $0.2\,\zsun$ respectively.}.

One possibility is a fast H{\sc i} gas flow, which is observed as the tidal interaction between
the LMC and the SMC \citep{Fukui2017a,Tsuge2019}.
Its velocity is as high as $50$--$100\,\kms$, comparable to the escape velocity of the LMC/SMC\@.
We can straightforwardly expect the coherent continuation of such a flow over a few 10 Myr timescale 
because it travels on a 10 kpc scale between the LMC and the SMC\@.
However, 
such a fast flow induces strong turbulence in molecular clouds,
which can also destroy dense structures (\cf, \citealt{Klein1994,Heitsch2006a}).
In addition, even for such efficient mass accumulation,
the thermal evolution from the WNM to molecular clouds 
still requires the cooling.
For example, 
albeit on $L=100$ pc scale with a coarser spatial resolution of 0.2 pc,
\cite{Maeda2021}
perform converging WNM flow simulations with $V_{\rm inflow}=100$ km s$^{-1}$ but 
without any CNM structure pre-existed in the WNM flow.
In the $0.2\,\zsun$ environment with the initial WNM density of $n_0 = 0.75\,\cmkk$,
they show that the molecular cloud formation takes 23 Myr.
The cloud mass with $n>10^4\,\cmkk$ is a few $10^5\,\msun$ at 23 Myr, which 
is consistent with Equation~\ref{eq:mcloud2}\footnote{In Equation~\ref{eq:mcloud1}, we count all the CNM mass
with $n>10^2\,\cmkk$. This therefore gives a typical maximum mass of formed molecular clouds.}.

Therefore, even in the fast flow environment, 
the pre-existence of CNM structure in the WNM is important
\citep[\cf,][]{Pringle2001,Inoue2009}.
This leads to another question ``how does the ISM form the CNM in the first place in low metallicity environments?''
As seen in our simulations, low-mass CNM clumps form if the mean magnetic field is parallel to compressive events
and this inflow can be as slow as $V_{\rm inflow}=20\,\kms$.
Therefore, some previous generations of a few 10 $\kms$ flow
along with the mean magnetic field
(such as a single supernova) are responsible to introduce CNM clumps in the WNM.
Such CNM pre-existence and subsequent compression due to shocks
presumably
determine the formation site, the mass, and the structure of molecular clouds.
Indeed, based on the recent ALMA observations toward N159E/W regions in the LMC,
\cite{Tokuda2022} indicate that a fast HI gas flow
with a multiple-scale substructure potentially explains the formation of filamentary
molecular clouds on various size scales.
Such a hierarchical structure may also determines the fractal nature of 
the young stellar structures in the LMC, suggested by the spatial clustering analysis 
of star clusters \citep[\eg,][]{Miller2022}.
In addition,
the similarity between the power-law spectrum of molecular cloud mass function 
(\eg, ${\rm d}n/{\rm d}m \propto m^{-\alpha}$ where $\alpha = 1.7$--$1.9$ in the LMC: \cite{Fukui2001,Ohno2023})
and that of CNM clump mass function in our simulation also indicates 
that the molecular cloud formation is pre-determined by the CNM structure
(Figure~\ref{fig:CNMMF}).
This implication needs further studies to confirm in the future.

\subsection{H$_2$ cooling}
\label{subsec:H2c}
The H$_2$ is an important coolant in the context of the primordial gas cooling.
Chemical paths to the H$_2$ formation exist even in the primordial gas,
especially that via the gas-phase interaction between H and H$^{-}$,
where the electron fraction impacts the abundance of H$^{-}$.
\cite{Susa1998} calculate the electron fraction in the postshock layer of supersonic ISM flows
and \cite{Susa2004} show that, with $V_{\rm inflow}=30\,\kms$, 
H$_2$ cooling is dominant over the metal line cooling in the postshock region 
even in $0.1\,\zsun$ without any UV background that dissociates the H$_2$.
This is not the case in our calculation where we employ $G_0$ = 1 for all the metallicity
by assuming ongoing active star formation as in the LMC/SMC\@. The metal line cooling is always dominant
in this setup.

The impact of the H$_2$ cooling on thermal instability is limited typically to $G_0 \leq 10^{-3}$ environments 
(\cite{Inoue2015}; see also \cite{Glover2014}).
It is left for future studies 
to investigate 
the dynamical formation and evolution of the multi-phase ISM 
driven by supersonic flows
in lower metallicity environment with UV radiation field of $G_0 <1$.

\subsection{Different assumptions: Dust-to-Gas ratio, electron fraction, and radiation field}
In this section, we introduce several previous studies 
to list how the adopted assumptions impact our results.

Firstly, 
many previous studies 
also assume that the dust-to-gas ratio is linearly scaled with the gas-phase metallicity in their fiducial models.
Meanwhile, based on the extragalactic observational indications \citep[\eg,][]{RemyRuyer2014},
there are also studies testing a broken power-law dependence. 
For example, $\mathcal{D} \propto Z^2$ in \cite{Glover2014} and $Z^3$ in in \cite{Bialy2019},
where $\mathcal{D}$ is the dust-to-gas ratio.
Such a superlinear decrease of the dust abundance at $\lesssim 0.2\,\zsun$ 
reduces the photoelectric heating rate relatively to the CII cooling rate.
The thermally stable states of the UNM and the CNM is cooler 
in this superlinear case\footnote{See also 
Figure~2 of \cite{Bialy2019} for the transition from the photoelectric heating dominated regime 
to the cosmic-ray heating dominated regime.}.
However in the range of $\geq 0.2\, \zsun$, they show that this difference has a limited impact 
on the CNM thermal equilibrium state (see also Figure~5 of \cite{Kim_jg2023}).
In addition, this superlinear relation of the dust-to-gas ratio does not significantly impact 
the overall dynamics in our simulations
because CII cooling at the shock-heated WNM right after the shock compression
provides the typical evolutionary timescale (as discussed in Appendix~\ref{subsec:tcool}).

Secondly, the gas-phase electron fraction changes the photoelectric heating efficiency. The photoelectric heating rate by FUV radiation can be estimated as $\Gamma_{\rm phot}(1.0\,\zsun) = 1.3 \times 10^{-24} \epsilon G_0 \, \mathrm{erg \, s^{-1}}$ where $\epsilon$ is the photoelectric heating efficiency as
\begin{eqnarray}
    \epsilon &=& \frac{4.9\times10^{-2}}{1+4.0\times10^{-3}(G_0 T^{1/2}/n_e\phi_{\rm PAH})^{0.73}} \nonumber \\
             && + \frac{3.7\times10^{-2}(T/10^4)^{0.7}}{1+2.0\times10^{-4}(G_0T^{1/2}/n_e\phi_{\rm PAH})} \,
             \label{eq:PErate}
\end{eqnarray}
(see Equation (43) in \citealt{Bakes1994}).
Here $n_e$ is the electron number density so that the electron fraction can be defined as $x_e = n_e/n_{\rm H}$, and the polycyclic aromatic hydrocarbon (PAH) reaction rate is $\phi_{\rm PAH}=0.5$ \citep{Wolfire2003}.
The $G_0 T^{1/2}/n_e$ dependence in the denominator describes the relative importance of the ionization against the recombination on the dust surface\footnote{See also Equations (19) and (20) in \cite{Wolfire1995} and Equations (32)--(34) in \cite{Bialy2019} for the exact form; \cf,\citealt{Bakes1994}.}. 
Our fiducial value, $\Gamma_{\rm phot}(1.0\,\zsun)=2.0 \times 10^{-26} \, \mathrm{erg \, s^{-1}}$ in Equation~\ref{eq:HeatingFunc}, corresponds to the typical condition of the thermally stable WNM at $1.0\, \cmkk$ with the $x_e \sim 0.02$ determined by cosmic ray ionization (\cf, Equation (12) of \citealt{Bialy2019}).
The photoelectric heating efficiency may deviate from this fiducial value right after the shock compression where the collisional ionization increases the electron fraction.
\cite{Koyama2000} performed a high-resolution simulation (albeit one-dimensional) of shock propagation into the WNM. 
The postshock shock-heated WNM volume reaches $n\sim5\,\cmkk$, $x_e\sim 0.1$, and $T~\sim 6400$ K, so that Equation~\ref{eq:PErate} expects that 
$\Gamma_{\rm phot}(1.0\,\zsun)=6 \times 10^{-26}\, \mathrm{erg \, s^{-1}}$. The factor of three difference form our fiducial value can expand the parameter space for the thermally stable WNM\@.
However note that a factor of few uncertainty also exists in the assumption on the PAH size distribution and shape \citep{Kim_jg2023}.
Also note that such high $x_e$ volume is limited to $10^{-1}$ pc scale from the shock front by the recombination (\ie, Figure 3 of \citealt{Koyama2000}).
Therefore we opt to use the fiducial constant photoelectric heating efficiency (and its linear dependence on the metallicity) for our comparison between metallicities in this article.

Lastly, the interstellar radiation field varies across space.
It is ideal to numerically investigate 
the 10pc-scale local radiation field,
\eg, zoom-in approach consistently coupled with galactic-scale star formation 
\citep[\eg,][]{Peters2017,Kim_CG2022,Kim_jg2023}
and this is left for future studies.

\subsection{Applicability of converging flow results to the ISM in reality}
\label{subsec:cfs}
Our results indicate that the solenoidal-mode-dominated turbulence is generally realized
in molecular clouds undergoing supersonic compressions in the metallicity range of 1.0--0.2 $\zsun$.
The density inhomogeneity (and/or velocity fluctuation) in the converging flows
can be larger (can exist) in reality, especially if the flows are already multiphase than the one-phase WNM \citep{Inoue2012,Iwasaki2019}.
Such a strong density/velocity inhomogeneity leads to stronger shock deformation
and induces stronger shear motion into the shock-compressed layer.
The stronger inhomogeneity in the inflow density/velocity induces stronger turbulence
\citep[\eg, the systematic survey by][]{Kobayashi2020},
but the turbulent speed is expected to be on the order of the WNM sound speed (\cf, Appendix~\ref{sec:TV}).

Expansion of multiple supernova remnants also induce cloud compressions 
as often observed in galactic scale numerical simulations \citep{Girichidis2016,Kim_CG2022}
and may explain the bubble features in nearby galaxies observed in JWST \citep{Watkins2023}.
Each compression event occurs at diverse angles at various timing 
and the generation of solenoidal mode turbulence locally occurs due to the deformed shock fronts.
It is likely that the compressive mode of turbulence does not become dominant 
unless multiple compressions from various angles occurs at a same timing with relatively uniform density
or unless the cloud itself becomes massive enough to self-gravitationally collapse.
Previous analytic studies also show that the solenoidal mode of turbulence can grow faster than the compressive mode
even in gravitationally contracting background \citep[\eg,][]{Higashi2021}.

\subsection{Implications to the cosmic star formation history}
\label{subsec:AppRan}
Our simulations show that 
physical properties of molecular clouds resemble if compared 
at the same time in the unit of the cooling time.
How about those in lower metallicity ranges, which are important to understand 
the overall cosmic star formation history beyond the Magellanic Clouds?
The metallicity dependence that we have shown comes primarily from 
the line coolings of [OI] ($63.2\, \mu$m) and [CII] ($157.7\, \mu$m).
This metallicity dependence in the cooling rate is expected to
continue down to $Z\sim 10^{-3}\zsun$, until which 
[OI] and [CII] are the dominant coolant to induce thermal instability.
In contrast, the main heating process changes from the photo-electric heating 
to the X-ray and cosmic ray heating at $\sim0.1\,\zsun$.
The heating rate in $<0.1\,\zsun$ becomes independent to the metallicity if we 
employ the same constant X-ray and cosmic ray heating rate. 
Under such conditions in $<0.1\,\zsun$,
\cite{Bialy2019} show that the density of the thermally stable CNM scales roughly as 
$n_{\rm CNM} \propto Z^{-1}$ (see their Equation~18).
This results in a roughly constant cooling time in $<0.1\,\zsun$.
In an even lower metallicity range of $Z\lesssim 10^{-3}\,\zsun$, 
the UV field strength impacts the growth of thermal instability
(due to $H_2$ dissociation)
and thus the comprehensive investigations remain 
by changing also the UV field strength for future studies.
Overall, we presume that, in the range of $\geq0.1\,\zsun$, physical properties of molecular clouds
is similar between metallicities 
at the same time measured in the unit of the cooling time,
and we need further investigation for in the range of $<0.1\,\zsun$.

Meanwhile, 
our results also indicate that 
the difference in the cloud formation condition (\eg, different $n_0$, $V_{\rm inflow}$ \etc) is important 
to form molecular clouds with different properties, which 
are required to comprehensively understand the cosmic star formation history.
As an example, recent ALMA observations started to spatially resolve galactic disk of 
star-forming galaxies at the so-called Cosmic Noon
at the redshifts of 2--4. They show the existence of molecular clouds in the gravitationally unstable galactic disks
and the clouds tend to have higher column density / higher star formation rate density
(\eg, $\Sigma_{\rm gas} > 100 \, \msun$ pc$^{-2}$ and $\Sigma_{\rm SFR}>1 \, \msun$ yr$^{-1}$ kpc$^{-2}$)
than those in the Milky Way \citep[\eg,][]{Tadaki2018}.
Such properties resemble those observed in nearby luminous infrared galaxies
\citep{Kennicutt2021}, who possibly experience galaxy mergers.
These indicate a possibility that some drastic mechanisms, such as galaxy mergers with high velocity,
form high column density molecular clouds that host active star formation at the Cosmic Noon.

In addition, 
even starting with the same cloud formation condition to form a similar molecular cloud across the metallicity,
the variation of the stellar initial mass function with the metallicity, if any,
may arise from subsequent fragmentation in collapse of the clouds and the circum-stellar disks.
We here remark that some previous authors investigated this phenomena with numerical simulations
using the same initial cloud properties
for all metallicities, such as \citealt{Bate2019,Chon2021a}.

The above discussions are just speculations
at this moment.
Further simulations with various
initial density, inflow velocity \etc, are needed
and left for future studies.

\section{Summary and Prospects}
\label{sec:summary}
To understand metallicity-dependence of molecular cloud formation
at its initial stages,
we perform and compare the MHD simulations of the WNM supersonic converging flows with $20\,\kms$
on 10 pc scales in $1.0$, $0.5$, and $0.2\,\zsun$ environments.
We impose the mean magnetic filed strength of 1 $\mu$G 
as representative strength in the metallicity range 
from the Milky Way to the Magellanic Clouds.
The field orientation is parallel to the supersonic flows,
which is a promising configuration for efficient molecular cloud formation.
The flow forms a shock-compressed layer sandwiched by two shock fronts,
within which thermal instability occurs to develop the multi-phase ISM\@.
We employ the 0.02 pc spatial resolution to resolve the thermal instability 
and turbulent structures.

We summarize our findings as follows:

\begin{enumerate}
    \item The development of the CNM structure in the shock-heated WNM 
          requires longer time in the lower metallicity environment where the typical $t_{\rm cool}$ is 
          almost inversely proportional to the metallicity.
          The CNM thermal states at different metallicities are similar if compared at 
          the same time measured in the unit of the cooling time,
          instead of the same physical time.
    \item The typical field strength of magnetic fields evolves gradually with $\langle B \rangle \propto n^{1/5}$
          up to $B\sim 11$ $\mu$G at $n\sim 10^3\,\cmkk$. 
          This is consistent with Zeeman measurements of low (column) density regions of molecular clouds
          in the Milky Way, and Faraday rotation measurements toward the ISM in the LMC and SMC.
    \item The postshock turbulent pressure balances against the inflow ram pressure (Panel (c) of Figure~\ref{fig:Turbulence}). 
          The turbulent velocity is slower in a lower metallicity environment (albeit the difference is within a factor of two),
          which is consistent with the line width of molecular clouds observed in the LMC/SMC.
    \item The velocity power spectrum follows the Kolmogorov's law if averaged over the entire volume of the shock-compressed layer,
          while two-point velocity correlation of the CNM volume alone exhibits the transition towards the Larson's law.
    \item At all metallicities after the $1.0\,t_{\rm cool}$,
          the solenoidal (compressive) mode of the turbulence in the shock-compressed layer 
          accounts for $>80$ percent ($<20$ percent, respectively) of the total turbulence power.
          This indicates that, even in a galactic-scale converging region,
          forming molecular clouds are always solenoidal-mode dominated.
          Therefore, a galactic-scale compressive motion is important for molecular cloud formation,
          but it does not immediately mean an enhancement of star formation efficiency by enhancing compressive motion in molecular clouds.
    \item The CNM clump mass function has a power-law distribution as ${\rm d}n/{\rm d}m \propto m^{-1.7}$,
          which can be explained by the thermal instability growth under the Kolmogorov turbulence background.
    \item These results suggest the common existence of hierarchical thermal and turbulent structure 
          in molecular cloud precursors in the $1.0$--$0.2\,\zsun$ range. 
          The WNM/UNM components occupy most of the volume with strong turbulence of $4$--$10\,\kms$.
          Meanwhile, the CNM component has the inter-clump velocity of $3$--$5\,\kms$ as well as the internal velocity dispersion
          of $\lesssim 1\,\kms$ within individual clumps (Figure~\ref{fig:sch_MC}).
    \item We expect that this similarity in molecular cloud properties across the metallicity at the same $t/t_{\rm cool}$ 
          continues to hold down to $Z\sim 10^{-3} \zsun$, because the dominant coolant are [OI] ($63.2\, \mu$m)
          and [CII] ($157.7\, \mu$m) until this metallicity.
          Meanwhile, this indicates that 
          some different cloud formation condition (\eg, different $n_0$, $V_{\rm inflow}$ due to galaxy mergers \etc) 
          is required to form molecular clouds with higher columnd density / higher star formation rate density,
          as observed in luminous infrared galaxies and star-forming galaxies at the Cosmic Noon.
    \item Our results show that, in the lower metallicity environment, 
          the longer physical time is required for the development of CNM structures out of the pure WNM\@.
          This indicates that,
          at the formation stage of molecular clouds out of the WNM in low-metallicity environments, 
          the pre-existence of CNM structure in the WNM volume
          controls the formation site and the mass of molecular clouds.
\end{enumerate}

Our calculation is still limited to the early phase of molecular cloud formation.
Investigating 
further compression in low metallicity environments is left for future studies.
In a $1.0\,\zsun$ environment, it is known that the efficiency of the molecular cloud formation depends 
on the inclination of the mean magnetic field against the inflow \citep[\eg,][]{Inoue2009,Iwasaki2019}.
We are planning to investigate similar dependence of the magnetic field geometry in our forthcoming article.

Collisions between flows with different metallicities is ubiquitous in the context of galaxy mergers, including the LMC-SMC tidal interaction.
It is interesting to investigate the spatial and temporal variation of the metallicity in the shock-compressed layer 
created by WNM flows with different metallicities, but is also left for future studies.

Studying cloud formation in extremely low metallicity environments down to $10^{-4}\,\zsun$ is also important
in revealing the initial condition of star formation in young galaxies.
For example, formation of close binaries of massive stars
in such environments is interesting as an origin of the massive binary black holes 
whose coalescence events are observed by gravitational waves \citep{Abbott2016_metallicity}.
Previous numerical studies in this context start with cosmological initial conditions \citep[\eg,][]{SafranekShrader2014a,SafranekShrader2014b},
or with an idealized model of a star forming core without its formation process \citep[\eg,][super-critical Bonnor-Ebert sphere of $10^4\,\msun$ 
with $n\sim 10^4\,\cmkk$]{Chon2021a},
or focus on the kpc-scale thermal instability without resolving core scales $\sim 0.1$ pc \citep[\eg,][]{Inoue2015}.
We are planning to reveal the formation and the internal structure of such star-forming clouds
as an extension of our studies in this article.
The dependence on radiation field strength and cosmic-ray/X-ray intensities are other important parameters in such conditions
\citep[\cf,][]{Susa2015},
but are also left for future studies (see also Section~\ref{subsec:H2c}).

\section*{ACKNOWLEDGMENTS}
We appreciate the reviewer for the careful reading and comments, which improved our draft.
Numerical computations were carried out on Cray XC50 
at Center for Computational Astrophysics, National Astronomical Observatory of Japan.
MINK (JP18J00508, JP20H04739, JP22K14080), 
KI       (JP19K03929, JP19H01938, JP21H00056), 
K.Tomida (JP16H05998, JP16K13786, JP17KK0091, JP21H04487),
TI       (JP18H05436, JP20H01944),
KO       (JP17H01102, JP17H06360, JP22H00149),
and 
K.Tokuda (JP21H00049, JP21K13962)
are supported by Grants-in-Aid from the Ministry of Education, Culture,
Sports, Science, and Technology of Japan. 
We appreciate Atsushi J. Nishizawa and Chiaki Hikage for helping our Fourier analysis,
and appreciate Kohei Kurahara for discussions on Faraday rotation measurements toward the Magellanic Clouds.
We are grateful to
Tomoaki Matsumoto, Hajime Susa, Sho Higashi, Gen Chiaki, Hajime Fukushima, Shu-ichiro Inutsuka,
Shinsuke Takasao, Tetsuo Hasegawa, Kengo Tachihara, and Jeong-Gyu Kim for fruitful comments.
We are grateful to Hiroki Nakatsugawa who contributed the early phase of this study
through his master thesis work.

\appendix
\section{Heating and Cooling Rate}
\label{sec:hcr}
We calculate the metallicity-dependent net cooling rate as
\begin{equation}
    \rho \myCooL(T,Z) = -\left(\frac{\rho}{\mymgas}\right) \Gamma(Z) +
    \left(\frac{\rho}{\mymgas}\right)^2 \Lambda(T,Z) \,,
\end{equation}
where $\mymgas = \mu_{\rm M} m_{\rm p}$.
The heating part consists of the photoelectric heating, X-ray heating, and cosmic-ray heating as
\begin{equation}
\begin{aligned}
    \Gamma(Z)            &= \Gamma_{\rm phot}(Z) + \Gamma_{\rm X} + \Gamma_{\rm CR} \,, \\
    \Gamma_{\rm phot}(Z) &= 2.0 \times 10^{-26} \left( \frac{Z}{1.0\,\zsun} \right) \, \mathrm{erg \, s^{-1}}    \,, \\
    \Gamma_{\rm X}       &= 2.0 \times 10^{-27} \, \mathrm{erg \, s^{-1}}    \,, \\
    \Gamma_{\rm CR}      &= 8.0 \times 10^{-28} \, \mathrm{erg \, s^{-1}}    \,. \\
\end{aligned}
\label{eq:HeatingFunc}
\end{equation}
Here, the photoelectric heating rate is proportional to the metallicity because it is dominated by dust grains \citep{Bakes1994}. 
On the other hand, X-ray and cosmic ray heating rates do not depend on the metallicity because they mostly heat hydrogen directly.

The cooling part consists of the cooling due to 
Ly$\alpha$, C$_{\rm II}$, He, C, O, N, Ne, Si, Fe, and Mg lines 
(see also \citealt{Koyama2000,Kobayashi2020}).
The functional form of the total cooling rate normalized 
in the unit of $\Gamma_{\rm phot}(1.0\,\zsun)$ is
\begin{equation}
\begin{aligned}
    &\frac{\Lambda(T,Z)}{\Gamma_{\rm phot}(1.0\,\zsun)} = \\
    &\begin{cases}
        10^7 \exp\left( \frac{-118400}{T+1000} \right) 
        + 1.4\times 10^{-2} \left(\frac{Z}{1.0\,\mathrm{Z}_{\odot}} \right) \sqrt{T} \exp \left( \frac{-92}{T} \right) \, \mathrm{cm^3}
        \hspace{3.5cm} \, (\mathrm{for} \, T\leq14,577 \,\mathrm{K})\,, \\ 
        5 \times 10^3 + 1.4\times 10^{-2} \left(\frac{Z}{1.0\,\mathrm{Z}_{\odot}} \right) \sqrt{T} \exp \left( \frac{-92}{T} \right) \, \mathrm{cm^3} 
        \hspace{3.4cm} (\mathrm{for} \, 14,577\,\mathrm{K}< T \leq 19,449 \,\mathrm{K})\,, \\ 
        \left(\frac{Z}{1.0\,\mathrm{Z}_{\odot}} \right) 
        \left[ 3.75 \times 10^4  
        \left( 1 -\tanh \left(\frac{T-2\times10^5}{2\times10^5}\right) \right) \exp \left(\frac{-5\times10^4}{T}\right) 
         +10^3 \exp \left( \frac{-5\times10^4}{T}\right) \right]\, \mathrm{cm^3}
        \hspace{0.1cm} (\mathrm{for} \, T > 19,449 \,\mathrm{K})\,. \\
    \end{cases}
    \label{eq:hcfunc}
\end{aligned}
\end{equation}
Here, the cooling rate by Ly$\alpha$ and He does not scale with the metallicity, while the other coolings due to the metals' lines are 
proportional to the metallicity.

\vspace{1cm}
\section{Cooling Time}
\subsection{The typical cooling time}
\label{subsec:tcool}
The cooling time, $P/(\gamma-1)/\rho\mathcal{L}$, varies locally with space and time,
depending on the local density/temperature variation. 
Nevertheless, the typical cooling time of this converging flow system can be estimated based on the 
typical state of the shock-heated WNM as follows.

To determine the representative thermal state, 
let us start with a simple estimation in a perfect one-dimensional shock compression.
Once the inflow WNM passes the shock front, 
the density in the downstream increases to
$n\simeq 4 n_0 = 2.3\,\cmkk$.
Meanwhile, the temperature also increase to
$T\simeq P_{\rm ram}/k_{\rm B}/(4n_0)=1.5\times10^4$ K,
but it rapidly decreases to $T\simeq T_0$
because of the efficient coolings by Ly$\alpha$ 
and heavy metals' lines. 
For example, the temperature decreases on the timescale of $2.0\times10^{-4}$ Myr,
$4.1\times10^{-4}$ Myr, and $1.0\times10^{-3}$ Myr at 1.0, 0.5, and 0.2 $\zsun$ 
based on the $T>14,577$ K regime of Equation~\ref{eq:hcfunc}
with $n=4n_0$ and $T= P_{\rm ram}/k_{\rm B}/(4n_0)$.
These timescales are shorter than the typical sound crossing time on a single numerical cell of 0.02 pc size 
with $T= P_{\rm ram}/k_{\rm B}/(4n_0)$,
which is $1.2\times 10^{-3}$ Myr.
Therefore, 
we need to perform simulations with a higher spatial resolution
to confirm whether this shock-heated WNM evolves isochorically or isobarically
during the return to $T\simeq T_0$.
In the following, let us assume that this rapid cooling leads to rather isochorical evolution
and employ $n= 4 n_0$ and $T=T_0$ as the representative initial density and temperature
to estimate the typical $t_{\rm cool}$ (see also the discussions 
in 
Section~\ref{subsec:deptcool}).
The corresponding pressure, $4 n_0 k_{\rm B} T_0$, is smaller than the ram pressure roughly by a factor of two 
($P/k_{\rm B} \simeq P_{\rm ram} /2k_{\rm B} = 1.75 \times 10^4 \, \mathrm{erg\,cm^{-3}}$).
This is consistent with most of the volumes in our simulations as shown in Figure~\ref{fig:TermStats}.
Such a thermal pressure smaller than the ram pressure is achieved also because
the turbulent pressure almost balances with the inflow ram pressure as shown in Figure~\ref{fig:Turbulence}.

Starting with $n= 4 n_0$ and $T=T_0$, the net cooling rate 
during the subsequent thermal evolution is mostly determined by the C$_{\rm II}$ cooling 
given by the first equation of Equation~\ref{eq:hcfunc}:
\begin{align}
    \rho\mathcal{L} &\simeq n^2 \Lambda(T,Z) \nonumber \\
                    &= n^2  \times 2.8 \times 10^{-28} (Z/\zsun) \sqrt{T} \exp(-92/T) \, \mathrm{erg \, s^{-1} \, cm^{-3}} \,.
\end{align}
Threfore, the cooling time can be estimated in general as
\begin{align}
    t_{\rm cool} &= \frac{P/(\gamma-1)}{n^2\Lambda(T,Z)} \label{eq:tcool1}  \\
                 &\simeq 2.99\,\mathrm{Myr} \, 
                       \left( \frac{P/k_{\rm B}}{10^4\,\mathrm{K\,\cmkk}}  \right) 
                       \left( \frac{n}{1\,\cmkk}                           \right)^{-2} 
                       \left( \frac{Z}{\zsun}                              \right)^{-1}
                       \left( \frac{T}{6400\,\mathrm{K}}                   \right)^{-1/2}
                  \exp \left( 1-\frac{T}{6400\,\mathrm{K}}                 \right) \label{eq:tcool2} 
\end{align}
With $n=4n_0$, $T=T_0$, and $P/k_{\rm B} = P_{\rm ram} /2k_{\rm B}$, 
the typical cooling time is
\begin{align}
    t_{\rm cool} &= 2.99\,\mathrm{Myr} \, 
                       \left( \frac{P_{\rm ram}/2k_{\rm B}}{10^4\,\mathrm{K\,\cmkk}}  \right) 
                       \left( \frac{4n_0}{1\,\cmkk}                           \right)^{-2} 
                       \left( \frac{Z}{\zsun}                              \right)^{-1}
                       \left( \frac{T_0}{6400\,\mathrm{K}}                   \right)^{-1/2}
                  \exp \left( 1-\frac{T_0}{6400\,\mathrm{K}}                 \right) \label{eq:tcool3}  \\
                 &\simeq 1.01 \, \mathrm{Myr} \left(\frac{Z}{\zsun}\right)^{-1} \,. \label{eq:tcool4} 
\end{align}
Equation~\ref{eq:tcool3} gives the typical value of $t_{\rm cool}$ as 
1.01 Myr at $1.0\,\zsun$,
2.03 Myr at $0.5\,\zsun$,
and 
5.11 Myr at $0.2\,\zsun$
as summarized in Table~\ref{table:ParamsMet}.
Equation~\ref{eq:tcool4} gives a rough metallicity-dependence.

\subsection{The dependence of the typical cooling time on the initial condition}
\label{subsec:deptcool}
In the current article, we focus on the metallicity dependence starting with the fixed initial condition.
Therefore, it is left for future studies to investigate the exact dependence of $t_{\rm cool}$ on the initial conditions other than the metallicity.
This requires parameter surveys by changing the mean density, the inflow velocity, the magnetic field strength, the inclination of the mean magnetic fields, the UV background, and so on.
The resultant impacts by all these changes are coupled each other because they modify the ram pressure, the turbulent pressure, and the following resultant shock-heated WNM state.

Nevertheless, as a first step, let us list several possible dependences to the two parameters
$n_0$ and $V_{\rm inflow}$. 
For simplicity, we here assume that the metallicity dependence comes only from the cooling/heating rate 
and is always independent from the choice of $n_0$ and $V_{\rm inflow}$.
The other conditions are as same as our simulation setup in this article
(\eg, the WNM inflow is parallel to the orientation of the mean magnetic field with $B=1$ $\mu$G, 
the injected WNM is on the thermally stable state \etc.).
These simplistic assumptions have to be investigated by further simulations, which is left for future studies.

Suppose that the postshock pressure, density, and temperature have the dependence as
$P\propto n_0^{\alpha} V_{\rm inflow}^{\beta}$, $n\propto n_0^{\gamma} V_{\rm inflow}^{\delta}$,
and $T = P/nk_{\rm B} \propto n_0^{\alpha/\gamma} V_{\rm inflow}^{\beta/\delta}$,
\begin{equation}
    t_{\rm cool} \propto n_0^{\frac{\alpha-3\beta}{2}} \, V_{\rm inflow}^{\frac{\beta-3\delta}{2}} \, Z^{-1}
\end{equation}
If we employ $n= 4 n_0$ and $T=T_0$ as the shock-heated WNM state right behind the shock
as we did in Section~\ref{subsec:tcool}, $\alpha=\gamma=1$ and $\beta=\delta=0$. The cooling time is 
\begin{equation}
    t_{\rm cool} \propto n_0^{-1} \, Z^{-1} \,. \label{eq:dep1}
\end{equation}
We employ Equation~\ref{eq:dep1} as our fiducial dependence to derive Equation~\ref{eq:mcloud2}.

Alternatively, most of the shock-heated WNM volume evolves close to $T=T_0$
even beyond $n>4n_0$ (Figure~\ref{fig:TermStats}), 
and thus we may consider that an isothermal shock approximation describes the overall postshock state.
In this case, $\alpha=\gamma=1$ and $\beta=\delta=2$. The cooling time is
\begin{equation}
    t_{\rm cool} \propto n_0^{-1} V_{\rm inflow}^{-2} \, Z^{-1} \,,
\end{equation}
accordingly.

We may also consider a simpler condition with $P=P_{\rm ram}$
and $T=T_0$, and $n=P_{\rm ram}/k_{\rm B}/T$.
This gives $\alpha=\beta=\delta=2$ and $\gamma=1$ and the cooling time is 
\begin{equation}
    t_{\rm cool} \propto n_0^{-1/2} V_{\rm inflow}^{-2} \, Z^{-1} \,, 
    \label{eq:OIA}
\end{equation}
However as shown in our simulations, 
such a perfect isothermal compression is limited to the central volume during the very early stage
when the shock compression is still close to a plane parallel geometry without significant deformation
(\cf, Panel (c) of Figure~\ref{fig:DB3view} and Panel (c) of Figure~\ref{fig:TermStats}).

Lastly, we would like to note that the above arguments assume 
the maximum pressure of the thermally stable state is independent of the metallicity.
The flow ram pressure in the $<0.1\,\zsun$ environments must be higher (e.g., with a higher $n_0$, a faster $V_{\rm inflow}$)
to successfully overcome this maximum pressure to enter the thermally unstable regime,
because the maximum pressure 
depends on the matallicity 
roughly as $\propto 1/Z$ \citep[see Figure~6 of][]{Bialy2019}.

\section{Turbulent Velocity}
\label{sec:TV}
Panel (a) of Figure~\ref{fig:Turbulence} shows that the turbulent velocity at earlier stages is much slower than the WNM sound speed, especially in lower metallicity environments (\eg, $\sim 2\,\kms$ in $0.2\,\zsun$ at $t \lesssim 0.5 t_{\rm cool}$). 
The shock front configuration controls this turbulence strength. As we discussed in the first two paragraphs of Section~\ref{subsec:thermal},
CNM forms slower in lower metallicity environments and most of the volume resides in the shock-heated WNM state for longer physical time. Its thermal pressure, comparable to the inflow ram pressure, 
keeps the shock fronts as the plane-parallel configuration.
(see Panels (a)--(c) of Figures~\ref{fig:DB3view} and~\ref{fig:TermStats}).
Such configuration decelerates the inflow more efficiently than deformed shock fronts, and the postshock volume becomes denser and less turbulent.
The physical state of the shock-heated WNM in this efficient deceleration regime 
can be estimated with the one-dimensional isothermal shock jump condition as
\begin{equation}
    \frac{v_1}{V_{\rm inflow}} 
    = \frac{n_0}{n_1} 
    = \frac{P_0}{P_1}
    = \mathcal{M}^{-2} \,.
\end{equation}
(See also Panel (c) of Figure~\ref{fig:TermStats} and the discussion after Equation~\ref{subsec:deptcool} for the validity of the one-dimensional isothermal approximation.)
Here $v_1$ is the postshock flow velocity, $n_1$ is the postshock density, $P_1$ is the postshock thermal pressure,
and $\mathcal{M}$ is the inflow Mach number. 
The setup of the injected WNM expects $v_1=2.0\,\kms$ and $n_1=5.8\,\cmkk$, which is seen especially in Panel (c) of Figure~\ref{fig:TermStats}.
This is not prominent in higher metallicity environments because the physical time  
of the plane-parallel shock configuration is shorter due to the efficient transition from the WNM to the CNM (\ie, the resultant interaction between CNM clumps and shock fronts 
induce more deformation of the shock fronts at earlier timing). 
In addition, in lower metallicity environments, the shock-heated WNM travels more distance from the shock front when they become the CNM\@.
Dense CNM clumps thus tend to form in the central region
in lower metallicity environments (Panel (e) of Figure~\ref{fig:DB3view}) 
and the turbulent development delays. This contributes to the difference between the metallicities in Panels (a) and (b) of Figure~\ref{fig:Turbulence} even after normalized with $t_{\rm cool}$ until the turbulence is fully developed at $t\sim 3 t_{\rm cool}$.

The turbulent velocity increases in time once the shock fronts start to deform. \cite{Kobayashi2020} performed a systematic survey to show that
the level of the shock deformation changes with the strength of the inflow density inhomogeneity. 
The turbulent velocity is, however, 
limited on the order of the WNM sound speed and it does not exceed $10\,\kms$.
This is a natural consequence of the oblique shocks.
As seen in our current simulation and \cite{Kobayashi2020}, 
the typical spatial scale of the shock deformation reaches the system size
(see also Figure 8 of \citealt{Kobayashi2020}).
Therefore, if we consider $\alpha \sim 45$ degree as the typical shock angle against the inflow,
the oblique isothermal shock jump conditions provide the postshock physical states as 
\begin{equation}
    \frac{v_1}{V_{\rm inflow}}  
    = \frac{n_0}{n_1}
    = \frac{P_0}{P_1}
    = \left(\mathcal{M} \sin\alpha \right)^{-2}   \sim 0.2\,.
    \label{eq:DeformPSdv}
\end{equation}
Here $v_1$ is the postshock velocity component normal to the shock front
and this gives $v_1\simeq 4.1\,\kms$. 
The angle is locally smaller ($\alpha< 45$ deg) to generate faster flow,
but $v_1>10\,\kms$ can be achieved when the inflow is parallel to the shock front with $\alpha\leq 27$ deg.
Such a closely parallel configuration does not occur on the entire area of the shock front 
so that the volume-averaged postshock velocity dispersion remains on the order of the WNM sound speed.


\bibliographystyle{aasjournal}
\bibliography{kobayashi_LowZ_MCformation_20230720}


\end{document}